\documentclass{aa}
\usepackage{graphicx}
\usepackage{txfonts}
\usepackage{url}
\usepackage{lineno}
\usepackage{soul}
\usepackage{amsmath}
\usepackage{xcolor}
\usepackage{mathtools}
\usepackage{natbib}
\bibpunct{(}{)}{;}{a}{}{,} 
\usepackage[version=4]{mhchem}
\usepackage{multirow}
\usepackage{placeins}
\usepackage[utf8]{inputenc}
\usepackage{hyperref}

\begin{document}

   \title{The effect of a biosphere on the habitable timespan of stagnant-lid planets and implications for the atmospheric spectrum}
    \titlerunning{The effect of life on habitability and the atmospheric spectrum}
    \authorrunning{D. Höning et al.}

 \author{Dennis Höning\inst{1}\thanks{DH and LC contributed equally to this paper.}
    \and
    Ludmila Carone\inst{2}\footnotemark[1]
    \and
    Philipp Baumeister\inst{3}
    \and
    Kathy L. Chubb\inst{2,4}
    \and
    John Lee Grenfell\inst{3}
    \and
    Kaustubh Hakim\inst{5,6}
    \and
    Nicolas Iro\inst{3}
    \and
    Benjamin Taysum\inst{3}
    \and
    Nicola Tosi\inst{3}    
    }
\institute{Potsdam-Institute for Climate Impact Research, Potsdam, Germany\\ \email{dennis.hoening@gmail.com}
    \and
    Space Research Institute, Austrian Academy of Sciences, Schmiedlstr. 6, A-8042 Graz, Austria
    \and
    Institute of Planetary Research, German Aerospace Center (DLR), Berlin, Germany
    \and
    Center for Exoplanet Science, University of St Andrews, North Haugh, St Andrews, UK
    \and
    Royal Observatory of Belgium, Ringlaan 3, 1180 Brussels, Belgium
    \and
    Institute of Astronomy, KU Leuven, Celestijnenlaan 200D, 3001 Leuven, Belgium}


         
 
\abstract
{Temperature-dependent biological productivity controls silicate weathering and thereby extends the potential habitable timespan of Earth. Models and theoretical considerations indicate that the runaway greenhouse on Earth-like exoplanets is generally accompanied by a dramatic increase in atmospheric H$_2$O and CO$_2$, which might be observed with the upcoming generation of space telescopes. If an active biosphere extends the habitable timespan of exoplanets similarly to Earth, observing the atmospheric spectra of exoplanets near the inner edge of the habitable zone could then give insights into whether the planet is inhabited. Here, we explore this idea for Earth-like stagnant-lid planets. We find that while for a reduced mantle, a surface biosphere extends the habitable timespan of the planet by about 1 Gyr, for more oxidising conditions, the biologically enhanced rate of weathering becomes increasingly compensated for by an increased supply rate of CO$_2$ to the atmosphere. Observationally, the resulting difference in atmospheric \ce{CO2} near the inner edge of the habitable zone is clearly distinguishable between biotic planets with active weathering and abiotic planets that have experienced a runaway greenhouse. For an efficient hydrological cycle, the increased bioproductivity also leads to a \ce{CH4} biosignature observable with JWST. As the planet becomes uninhabitable, the H$_2$O infrared absorption bands dominate, but the 4.3-micron CO$_2$ band remains a clear window into the \ce{CO2} abundances. In summary, while the effect of life on the carbonate-silicate cycle leaves a record in the atmospheric spectrum of Earth-like stagnant-lid planets, future work is needed especially to determine the tectonic state and composition of exoplanets and to push forward the development of the next generation of space telescopes.
}

\maketitle

\section{Introduction}
\label{sec:intro}
The search for life on planets beyond our Solar System is receiving a boost with the new and upcoming generation of space telescopes enabling the characterisation of planetary atmospheres \citep[e.g.][]{gialluca2021characterizing,fauchez2019impact,Lin2022}. A detection of a chemical disequilibrium -- such as oxygen together with methane -- would indicate that oxygen is replenished continuously, and oxygenic photosynthesis is the most efficient way to maintain this disequilibrium \citep[e.g.][]{schwieterman2018exoplanet,krissansen2022understanding}.

A detailed characterisation of planetary atmospheres, however, requires expensive, long-term space telescope observations. Given the vast number of exoplanets and the limitations in observational time and cost, it is essential to prioritise the study and observations of planets that are most likely candidates to support the emergence and maintenance of life.

To first order, planetary habitability can be assessed using the concept of the habitable zone, which is the distance of a planet to the star that would allow for liquid water on the planetary surface. Besides the stellar luminosity and the orbital distance of the planet, greenhouse gases in the planetary atmosphere, in particular CO$_2$, determine the planetary surface temperature and thereby its potential habitability \citep[e.g.][]{kopparapu2013habitable,ramirez2018more}.

The concentration of atmospheric CO$_2$ is controlled by numerous complex processes involving the interior, the surface, and the atmosphere \citep[e.g.][]{foley2015role,kruijver2021,oosterloo2021}. Depending on the composition and thermal state of the mantle, CO$_2$ is released into the atmosphere via volcanism. If liquid surface water and an active hydrological cycle are present on the planet, CO$_2$ reacts with rainwater to form carbonic acid, which can dissolve silicate rocks. Weathering products are washed into the ocean, where calcium carbonate is precipitated. If the planet possesses Earth-like plate tectonics, carbonates are recycled back into the deep interior at subduction zones \citep[e.g.][]{sleep2001,kasting2003evolution,Catling_Kasting2017}.

On a planet without plate tectonics (i.e. on a stagnant-lid planet), carbonates are not subducted but instead accumulate on the seafloor where they are eventually buried by new lava flows, as long as the planet can sustain active volcanism. Therefore, layers of carbonated rocks gradually migrate downwards, heating up until they become unstable. Through this decarbonation reaction, CO$_2$ is released back into the atmosphere, thereby closing the carbonate-silicate cycle on a timescale shorter than for planets with active plate tectonics \citep{Foley:2018,honing2019carbon,baumeister2023redox}.

In our Solar System, Earth is the only planet with plate tectonics. Theory and simulations indeed indicate that a stagnant lid is a more likely outcome of mantle convection \citep[e.g.][]{ogawa1991,davaille1993,moresi1995}. In fact, in the absence of efficient weakening mechanisms, the strong temperature dependence of the viscosity of mantle rocks naturally leads to the formation of a highly viscous, immobile, and conductive lid, extending from the surface down to the base of the thermal lithosphere \citep{Schubert:2001}. In this paper, we specifically explore the habitability of stagnant-lid planets.

Outgassed volatiles originate from surface lavas that are the product of partial melting of the upper mantle. The mantle composition thus plays an important role in the buildup and long-term evolution of the atmosphere and in turn on habitability. In particular, the mantle redox state controls whether reduced species (such as H$_2$, CO, or CH$_4$) or oxidised species with a strong greenhouse potential (such as H$_2$O and CO$_2$) are ultimately released into the atmosphere \cite[see][for a recent review]{gaillard2021}.

Deriving the interior composition and redox state of exoplanets is difficult and fraught with large uncertainties. In principle, rock-building elemental abundances of the host star could be used as proxies for the interior composition \citep[e.g.][]{dorn2015,brugger2017,unterborn2019}, but the link between the two is not straightforward \citep{plotnykov2020,schulze2021}. The spectroscopic study of `polluted' white dwarfs, which contain traces of accreting rocky bodies previously orbiting around them, suggests that the oxidation state of these bodies is similar to that of moderately reduced bodies of the Solar System such as Mars, although the discussion remains open about this \citep{doyle2019}. For this work, we treated the redox state of the mantle as a free parameter and followed the approach of redox melting that \citet{grott2011volcanic} applied to Mars and \citet{tosi2017habitability} and \citet{godolt2019habitability} did to exoplanets, focusing mainly on volcanic outgassing of CO$_2$. 

Besides tectonics and interior composition, a key factor that controls the climate and the habitability -- at least on Earth -- is its biosphere. Land vegetation enhances continental weathering, for example as roots of trees break rocks and thereby amplify the weathereable surface area, or with lichens that provide a continuously humid environment \citep[e.g.][]{berner1992weathering,schwartzman1989biotic,schwartzman1991biotic}. In addition, microbes and fungi produce acids that enhance silicate weathering. Since the biological productivity, and in turn the biological enhancement of weathering, increase with temperature and CO$_2$, negative feedback to temperature oscillations emerges: increasing temperature amplifies the biological enhancement of weathering by which CO$_2$ is effectively removed from the atmosphere, causing a decrease in the global mean surface temperature. This feedback process is particularly important as it can dampen the increase in surface temperature as stellar luminosity rises with time and can therefore potentially extend the habitable timespan of the planet \citep{lenton:2001}. Marine organisms also provide negative feedback to climate oscillations. However, this effect is mainly important on sub-million-year timescales \citep{honing2020impact}. Because our models are applicable on longer timescales (\textgreater 100 Myr), the effect of marine organisms is neglected in the present study.

For stagnant-lid planets, the effect of a surface biosphere on the surface habitability is not trivial. Even though biological enhancement of weathering provides negative feedback to rising temperatures by enhancing the drawdown of CO$_2$, carbonates are precipitated at a higher rate with an active biosphere, which enhances the carbon concentration of the crust. As the CO$_2$ is subsequently released into the atmosphere when carbonates become unstable, biological enhancement of weathering indirectly raises the rate at which CO$_2$ is released into the atmosphere. In this paper, we explore these opposing effects.

A key consequence of weathering is that most of the near-surface CO$_2$ is trapped in the form of carbonates, leaving only a small fraction of CO$_2$ in the atmosphere for planets inside the habitable zone. By the time the moist or runaway greenhouse limits are reached, oceans evaporate and weathering becomes increasingly inefficient or eventually ceases \citep{Kasting1988}. Since CO$_2$ will continue to be released into the atmosphere through mantle degassing, atmospheric CO$_2$ will rapidly increase. Such a bimodal distribution of CO$_2$ for planets on both sides of the habitable zone might be statistically identified by measuring CO$_2$ levels \citep{bean2017,graham2020,schlecker2023}. Measurements of CO$_2$ levels and Sun-like insolation with future mission concepts for at least 83 planets in the habitable zone have been suggested to determine if weathering is a universal phenomenon \citep{lehmer2020carbonate}.

For stagnant-lid planets, the rise of atmospheric CO$_2$ after runaway greenhouse is particularly strong, since mantle degassing is accompanied by self-accelerating crustal decarbonation: with increasing surface temperature, the depth at which decarbonation occurs gradually moves upwards, releasing even more CO$_2$ into the atmosphere. \cite{honing2021early} find that the atmospheric CO$_2$ partial pressure increases by $\approx$2 orders of magnitude within $\approx$100 Myr. We note that only a significant surface temperature rise -- such as during runaway greenhouse -- noticeably affects the decarbonation depth; in contrast, as long as the planet's surface temperature remains within the habitable range, the decarbonation depth is almost solely controlled by the mantle temperature \citep{honing2021early}.

If an active surface biosphere enhances the habitable timespan of exoplanets, it would shift the moment in time -- or the critical stellar distance -- at which the atmospheric CO$_2$ dramatically increases. Planets of a similar composition, age, and incident insolation could then either have small atmospheric CO$_2$ content -- if they are inhabited, with an active biosphere that enhances continental weathering -- or have a high atmospheric CO$_2$ content if the planet is abiotic and underwent a runaway greenhouse in its past. Determining the atmospheric CO$_2$ concentration of exoplanets near the inner edge of the habitable zone could then not only reveal whether or not the planet is habitable, but could also yield insights into a potentially active surface biosphere.

The goals of this paper are threefold. First, we aim to assess whether, and to which extent, a biosphere extends the habitable timespan of stagnant-lid planets. Second, we explore the differences in atmospheric CO$_2$ and CH$_4$ levels between inhabited (biotic) planets and uninhabited (abiotic) planets that have experienced a runaway greenhouse due to a shorter habitable timespan. Finally, we evaluate the observational potential for detecting these differences by modelling transmission measurements from the James Webb Space Telescope (JWST).

\section{Model}
\label{sec:model}

We employed a coupled interior-atmosphere evolution model for stagnant-lid planets \citep{grott2011volcanic,tosi2017habitability,godolt2019habitability,honing2021early,baumeister2023redox}. The interior thermal evolution is based on a standard parameterisation of convective heat transport based on solving the energy-conservation equations of the core, mantle, and stagnant lid. The convective heat flux is derived from boundary layer theory \citep{Schubert:2001} and the viscosity is strongly temperature-dependent \citep{grasset1998,choblet2000}.

Whenever the upper mantle temperature exceeds the solidus of dry peridotite \citep{katz2003}, the accompanying volume of melt is calculated assuming a linear increase in melt fraction between solidus and liquidus and accounting for latent heat. The melt is enriched in heat-producing elements and the residual mantle is depleted in those according to a partition coefficient. The produced melt is the only source of new basaltic crust and therefore essential for weathering. The melt is also enriched in CO$_2$ according to the parameterisation introduced by \citet{grott2011volcanic}, which assumes that carbon can be supplied to the melt as long as the mantle is sufficiently hot to undergo partial melting and in a concentration that only depends on the mantle oxygen fugacity. The generated melt volume is extracted instantaneously to the surface where CO$_2$ in excess of its solubility in basaltic melts is released into the atmosphere. The CO$_2$ outgassing rate therefore increases with the mantle oxygen fugacity.

We based the calculation of silicate weathering on stagnant-lid planets $F_w$ on \cite{honing2019carbon} but extended it to account for both the temperature- and CO$_2$ dependence of weathering:
\begin{linenomath}
\begin{equation}
    F_{w}^*=\omega^* B^* (p_{CO_2}^{*})^\alpha\exp \left(\frac{E_a}{R}\left(\frac{1}{T_{s,E}}-\frac{1}{T_s}\right)\right), 
\label{eq:F_s}
\end{equation}
\end{linenomath}
where $\omega^*$ is the relative weatherability (see below), $B$ is the biological enhancement of weathering, $p_{CO_2}$ is the partial pressure of CO$_2$ in the atmosphere, $T_s$ is the surface temperature, $\alpha$ is a constant, $E_a$ is the activation energy and $R$ is the gas constant. Here and in the following, the index $E$ denotes present-day Earth variables and the asterisk denotes variables scaled with their present-day Earth value, that is, $F_w^*=F_w/F_{w,E}$. Constants and free parameters are given in tables \ref{tab1} and \ref{tabp}, respectively.

\begin{table}
\caption{Constants used in the model.}
\label{tab1}
\centering
\begin{tabular}{p{0.7cm} p{3.7cm} l}
\hline
Para-meter & Description & Value \\
\hline
$\alpha$ & Constant, Eq. \ref{eq:F_s} & 0.33 $^{(1)}$\\
$E_a$ & Eff. activation energy for silicate weathering & $3.8\cdot 10^{4}$ J mol$^{-1}$ $^{(2)}$\\
$M_O$ & Ocean mass & $1.35\cdot 10^{21}$ kg $^{(1)}$\\
$a_{min}$ & Min. atm. CO$_2$ for plant growth & 10 ppm $^{(3)}$ \\
$B_{T,E}$ & Constant & 0.8352 ppm $^{(4)}$ \\
$a_{1/2}$ & Constant & 181 ppm $^{(4)}$ \\
$F_{sf,E}$ & Pre-industrial seafloor-weathering rate & $1.75\cdot 10^{12}$ mol yr$^{-1}$ $^{(5)}$\\
$T_{s,E}$ & Pre-ind. surf. temperature & 287 K\\
$p_{CO_2,E}$ & Pre-ind. atm. CO$_2$ & 280 ppm\\
$\xi$ & Pre-ind. fraction of abiotic weathering & 0.25 $^{(6)}$\\
$f_{degas}$ & Frac. temp. crustal carbon reservoir that degasses & 0.02 $^{(7)}$ \\
\hline
\end{tabular}
\flushleft{Sources: $^{(1)}$ \cite{Krissansen-Totton:2017}, $^{(2)}$ \cite{palandri2004compilation}, $^{(3)}$ \cite{bergman:2004}, $^{(4)}$ \cite{honing2020impact}, $^{(5)}$ \cite{mills:2014}, $^{(6)}$ upper value considered by \cite{lenton:2018}, $^{(7)}$ \cite{honing2019carbon}.}
\end{table}

\begin{table}
\caption{Free parameters used in the model.}
\label{tabp}
\centering
\begin{tabular}{p{0.8cm} p{2.5cm} l}
\hline
Para-meter & Description & Value \\
\hline
$d$ & Stellar distance & 0.7 AU -- 1.0 AU \\
$f_{O_2}$ & Mantle oxygen fugacity & IW-0.2 (IW-0.4, IW+0)\\
$T_{m,E}$ & Ini. mantle temp. & 1900 K (1850 K, 1950 K)\\
$\eta_{ref}$ & Ref. mantle visc. & $10^{21}$ Pa s ($10^{20}$ Pa s, $10^{22}$ Pa s)\\
\hline
\end{tabular}
\end{table}

As in \cite{honing2019carbon}, we scaled the relative weatherability $\omega^*$ on stagnant-lid planets to seafloor weathering on present-day Earth, since both rates directly depend on crustal production:
\begin{linenomath}
\begin{equation}
   \omega^*=\frac{X_E \xi_E}{f_E}\left(\frac{\mathrm{d}M_{cr}}{\mathrm{d}t}\right),
    \label{eq:weathersl}
\end{equation}
\end{linenomath}
where $\frac{\mathrm{d}M_{cr}}{\mathrm{d}t}$ is the crustal production rate and $X_E$, $\xi_E$, and $f_E$ are the present-day Earth values of mid-ocean ridge CO$_2$ concentration in the melt, fraction of seafloor weathering, and fraction of buried carbonates that enter the mantle. We note that the scaling to seafloor weathering solely serves the purpose of obtaining a relationship between the weatherability and rate at which fresh basaltic crust is produced; mid-ocean ridges producing fresh seafloor do not exist on stagnant-lid planets.

Modelling the biological enhancement of weathering (factor $B$ in Eq. \ref{eq:F_s}), we restricted ourselves to land plants as present on the modern Earth. Therefore, the model is strictly valid only for planets with a biosphere similar to that having emerged on Earth only 400-500 million years ago. We further assumed that a constant fraction $\xi$ of continental weathering is not affected by bioactivity whereas the rest (1-$\xi$) depends on plant productivity \citep{honing2020impact,lenton:2001,caldeira:1992}. Following \cite{honing2020impact}, we modelled plant productivity as a function of the surface temperature and atmospheric CO$_2$
\begin{linenomath}
\begin{equation}
    B^*=\xi+(1-\xi) \cdot B_T^* \cdot B_{CO_2}^*,
\label{eq:biow1}
\end{equation}
\end{linenomath}
where $B_T^*$ and $B_{CO_2}^*$ are the temperature- and CO$_2$-dependent terms of plant productivity. The temperature dependence follows a parabolic function of temperature
\begin{linenomath}
\begin{equation}
    B_T^*=B_{T,E}^{-1}\left(1-\left(\frac{T_s^{\left(^\circ\textrm{C}\right)}-25^\circ \textrm{C}}{25^\circ \textrm{C}}\right)^2\right),
\label{eq:biow2}
\end{equation}
\end{linenomath}
and the CO$_2$-dependence follows a Michaelis-Menton function
\begin{linenomath}
\begin{equation}
    B_{CO_2}^*=\frac{(P_{CO_2}-a_{min})(P_{CO_2,E}-a_{min}+a_{1/2})}{(P_{CO_2}-a_{min}+a_{1/2})(P_{CO_2,E}-a_{min})},
\label{eq:biow4}
\end{equation}
\end{linenomath}
where $B_{T,E}$, $a_{min}$ and $a_{1/2}$ are constants \cite[see][for details]{honing2020impact}. Specifically, we modelled bioproductivity to increase with temperature up to 25$^\circ$C and then to decrease at higher temperatures, so that the negative feedback is only effective until that specific surface temperature.

Carbonates were assumed to accumulate on the seafloor and be buried by new lava flows. Buried carbonated crust was tracked downwards until decarbonation occurs \citep{honing2019carbon,baumeister2023redox}. Decarbonation was assumed to occur through the breakdown of dolomite \citep{Foley:2018}. For more details, the reader is referred to \cite{honing2019carbon,honing2021early}.

Water vapour and CO$_2$ were considered as the only greenhouse gases and a radiative grey atmosphere model was used to calculate greenhouse heating \citep{Catling_Kasting2017}. Water outgassing from the mantle was neglected and the surface water budget was assumed to resemble Earth's. In synthesis, the surface temperature $T_s$ was calculated as a function of the equilibrium temperature $T_{eq}$ and of the optical depth of the atmosphere in the infrared $\tau$:
\begin{equation}
    T_s^4 = T_{eq}^4 \left(1+\frac{3\tau}{4}\right),
\end{equation}
where
\begin{equation}
    T_{eq}^4 = \frac{(1-A)S_{sun}}{4\sigma},
\end{equation}
where $S_{sun}$ is the insolation at the top of the atmosphere, $\sigma$ is the Stefan-Boltzmann constant, and $A$ is the albedo. The optical depth $\tau$ simply considers the additive contribution of the optical depths of H$_2$O and CO$_2$ expressed in terms of the corresponding absorption coefficients \cite[see][for details]{honing2021early,baumeister2023redox}.

\cite{honing2021early} compares the global mean temperature throughout the habitable period obtained by this model with the outcome of the 3-D global circulation model ROCKE-3D \citep{way2017,way2020} and obtain good agreement. We assumed that the modelled exoplanet has Earth-like size and mass, an albedo of 0.35, and orbits a Sun-like star accounting for stellar evolution. For application to a TRAPPIST-1-like star, we assumed that TRAPPIST-1e is the planet that is best suited for the scenarios explored in this work as it also orbits at the inner edge of its host star's habitable zone. 

Using our baseline model, we further explored exemplary snapshots in time of our sample planets with an atmosphere model. For the habitable, biotic cases (with will be later referred to as scenarios 1 and 2, see Table \ref{tab:2}), we used the fully self-consistent atmosphere-chemistry model \texttt{1D TERRA} \citep{Wunderlich20}, which takes the surface pressures of \ce{CO2} and \ce{H2O} from the geophysical model as well as 0.2~bar \ce{O2} and 1~bar \ce{N2} as input parameters. The model \texttt{1D TERRA} also assumes an active hydrological cycle, where we differentiated between a `dry' or `moist' composition. The effect of bioproductivity on the source fluxes in \texttt{1D TERRA}, as described in \cite[][Table~4]{Wunderlich20}, was established by taking into account the CO$_2$- and temperature-dependence of bioproductivity (Eq. \ref{eq:biow2} and \ref{eq:biow4}), resulting in biogenic fluxes of constituents such as CH$_4$ relative to present-day Earth by a factor of 1.905 and 1.805 for scenarios 1 and 2, respectively. For more details on the model, the reader is referred to section~\ref{sec:results2} and Appendix~\ref{sec: TERRA}.

For the abiotic cases (scenarios 3 and 4), we enter an atmosphere regime with uncertainties in atmospheric escape for a runaway-greenhouse scenario, in particular around M dwarfs \citep{boukrouche2021,Owen2019,Tiang2015,Gronoff2020} and in the role of photochemistry in dense \ce{CO2}-dominated atmospheres like Venus \citep{Stolzenbach2023,Wilson2024,Petkowski2024}. For these scenarios, we used a simpler model than for the biotic scenarios (Appendix~\ref{sec: simple atmo}). We assumed that water vapour cannot condense out of the atmosphere and further assumed strong vertical mixing. In these steam-dominated atmospheres, the atmospheric content of both \ce{CO2} and \ce{H2O} was determined by the geophysical model and assumed to be constant to first order, which is a typical assumption for runaway-greenhouse atmospheres \citep[e.g.][]{boukrouche2021,Lichtenberg2021}. We again assumed 1~bar \ce{N2} to which the respective partial pressures of \ce{H2O} and \ce{CO2} are added. We then applied the multispecies pseudoadiabat prescription by \cite{graham2021} to construct pressure--temperature profiles. For the `steam atmospheres', water condensation is in equilibrium with evaporation and rainout was neglected such that the strong latent heat effect of water dominates the temperature of the upper atmosphere.

Further, we explored a `desiccated' scenario, where we assumed to first order that the steam scenario 4 results in complete desiccation, which we mimicked by removing by \ce{H2O} completely while the resulting \ce{CO2} surface pressures is about 10\% of that of current Venus (Table~\ref{tab:2}). Readers can refer to Appendix~\ref{sec: simple atmo} for a more detailed overview. We further note that the complete removal of about one terrestrial ocean equivalent of water (270~bar atmospheric pressures) in a steam-dominated atmosphere can be assumed to be a fairly efficient process occurring within a time interval shorter than 100~million years \citep[e.g.][]{Abe2011,Barth21}.

\section{Results}

The results section is structured as follows. In section \ref{sec:results1}, we explore the effect of a biosphere on the partial pressure of atmospheric CO$_2$ focusing on the transition from a habitable to an uninhabitable climate. We start with our baseline model for which we used a mantle oxidation state of IW-0.2 (that is, a mantle oxidation state of 0.2 log units below the iron-w\"ustite-buffer), an initial mantle temperature of 1900 K, and a reference mantle viscosity of 10$^{21}$ Pa s. We chose this parameter combination for our baseline model for illustrative purposes as it results in a habitable period between 1.5 and 5 Gyr depending on the stellar distance, which we vary between 0.7 and 1 AU. Following this, we explore the sensitivity of our results to these parameters. In section \ref{sec:results2}, we then use our baseline model as an example to study the observational signatures associated with the transition towards uninhabitable conditions.

\subsection{Impact of bioproductivity on atmospheric \ce{CO2}}
\label{sec:results1}
\begin{figure*}
    \centering
    \includegraphics[width=\textwidth]{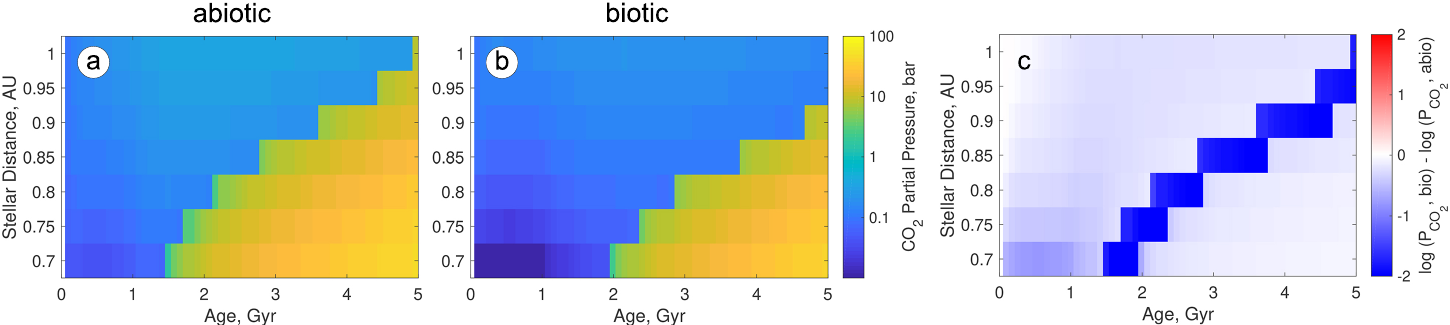}
    \caption
    {Resulting atmospheric CO$_2$ from our interior-atmosphere model. We show model results of an abiotic planet (a), a biotic planet (b), and the difference between them (c). The planet parameters are $f_{O_2}$=IW-0.2, $T_0$=1900 K, and $\eta_{ref}=10^{21}$ Pa s.}
    \label{fig:co2}
\end{figure*}

The bimodal distribution of atmospheric CO$_2$ (Fig. \ref{fig:co2}) indicates a transition from habitable surface conditions to uninhabitable conditions after the runaway greenhouse. The closer the planet is to the star, the earlier the runaway greenhouse sets in. For our baseline model, the habitable period of a lifeless (abiotic) planet (Fig. \ref{fig:co2}a) is shorter than for an inhabited (biotic) planet (Fig. \ref{fig:co2}b). The reason is the biologically enhanced rate of weathering, removing CO$_2$ at a rate much higher than for the lifeless planet. Fig. \ref{fig:co2}c depicts the difference between the two planets, indicating an extension of the habitable period by up to 1 Gyr if life is present (dark blue region in Fig. \ref{fig:co2}c).

\begin{figure*}[!ht]
    \centering
    \includegraphics[width=\textwidth]{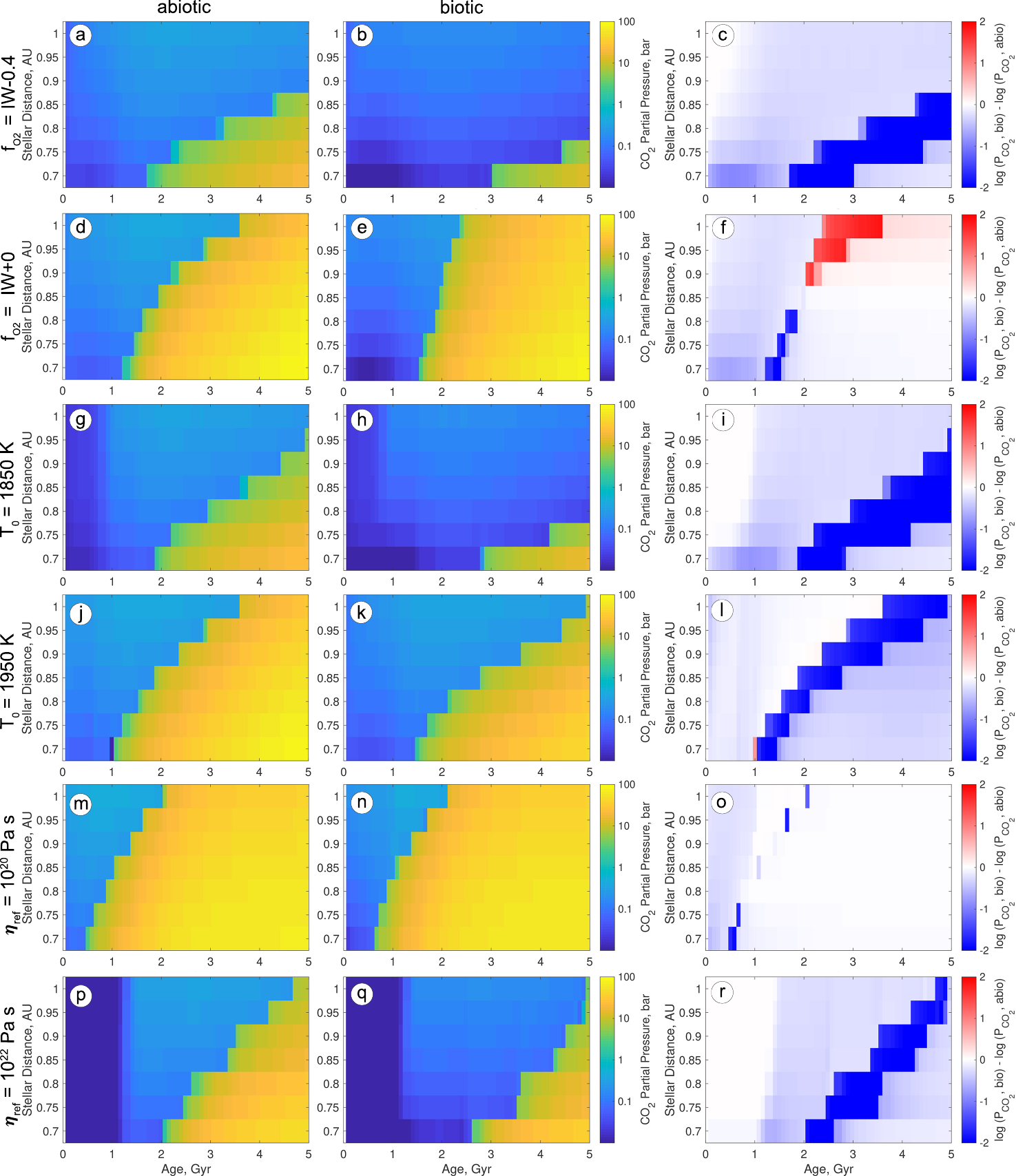}
    \caption
    {Atmospheric CO$_2$ of an abiotic planet (left column), a biotic planet (centre column), and the difference between the two (right column). We tested different oxygen fugacities $f_{O2}$ (a--f), initial mantle temperatures $T_0$ (g--l), and reference mantle viscosities $\eta_{ref}$ (g--l) keeping the respective other parameters equal to the baseline model ($f_{O2}$ = IW-0.2, $T_0$ = 1900 K, $\eta_{ref}$ = 10$^{21}$ Pa s).}
    \label{fig:parameterstudy}
\end{figure*}

In addition to the stellar distance, planet age, and a potential biosphere, other planet-specific parameters can control the atmospheric CO$_2$ and therefore the habitable timespan of stagnant-lid planets. In particular, the three parameters -- mantle oxidation state, initial mantle temperature, and reference mantle viscosity -- have a strong impact on the CO$_2$ outgassing rate throughout the evolution of stagnant-lid planets and therefore shift the moment in time the planet becomes uninhabitable. In Fig. \ref{fig:parameterstudy}, we systematically show the impact of the mantle oxygen fugacity (a--f), initial mantle temperature (g--l), and reference mantle viscosity (m--r).

For more reducing conditions ($f_{\rm O_2}$=IW-0.4, Fig. \ref{fig:parameterstudy}, panels a--c), the runaway greenhouse sets in later than for our baseline model ($f_{\rm O_2}$=IW-0.2, Fig. \ref{fig:co2}). This is due to the fact that the rate of CO$_2$ outgassing roughly changes proportionally to variations of the oxygen fugacity with respect to the IW-buffer \citep{grott2011volcanic,tosi2017habitability}. We also find that the impact of a biosphere is larger in this case, extending the habitable timespan by up to 2 Gyr. On the contrary, for more oxidising conditions ($f_{\rm O_2}$=IW+0, Fig. \ref{fig:parameterstudy}, panels d--f), the effect of biologically enhanced weathering can be overcompensated by the enhanced crustal carbon concentration, which is an indirect effect of the enhanced weathering rate, leading to an enhanced rate of crustal decarbonation. As a result, the habitable period could become shorter altogether. We note that this result is exclusively observed for large orbital distances (here: $\geq$0.9 AU; red region in Fig. \ref{fig:parameterstudy}f), which allow for a substantial increase in the crustal carbon concentration, and therefore of the rate of crustal decarbonation with time.

Another important parameter for stagnant-lid planets is the initial mantle temperature \citep[e.g.][]{tosi2017habitability,noack2017,dorn2018,honing2019carbon,honing2021early}. In Fig. \ref{fig:parameterstudy}g--l, we explore the effect of initial mantle temperatures of 1850~K and 1950~K (baseline model: 1900 K). A high initial mantle temperature causes rapid early outgassing, implying a higher level of atmospheric CO$_2$ during the subsequent evolution. Ultimately, this affects the inner boundary of the habitable zone: the higher the initial mantle temperature, the earlier the runaway greenhouse sets in. We find that the biosphere in both cases has a substantial effect on the habitable timespan, particularly striking for the lower initial mantle temperature.

Finally, another controlling parameter in planetary evolution models is the mantle viscosity, which depends on the planet's composition. Assuming a pressure- and temperature dependent viscosity following \cite{grasset1998} and \cite{choblet2000}, for our baseline model we set the reference viscosity to $\eta_{ref}=10^{21}$ Pa s. In Fig. \ref{fig:parameterstudy}g--l, we additionally tested reference mantle viscosities of 10$^{20}$ and $10^{22}$ Pa s. We find that a lower reference mantle viscosity substantially shortens the habitable timespan. This is a natural outcome of the more vigorous mantle convection, implying higher rates of outgassing particularly during the early evolution. Even a stellar distance of 1 AU does only allow for a habitable timespan of 2 Gyr. CO$_2$ accumulates rapidly and almost undiminished in the atmosphere, quickly making the surface conditions uninhabitable. We also observe that the impact of the biosphere is very limited in this case (extending the habitable timespan by only about 100 Myr). This is due to the fast circulation time of carbonates until they reach the decarbonation depth, which again is a result of the rapid mantle convection.

\begin{figure*}[!ht]
    \centering
    \includegraphics[width=0.75\textwidth]{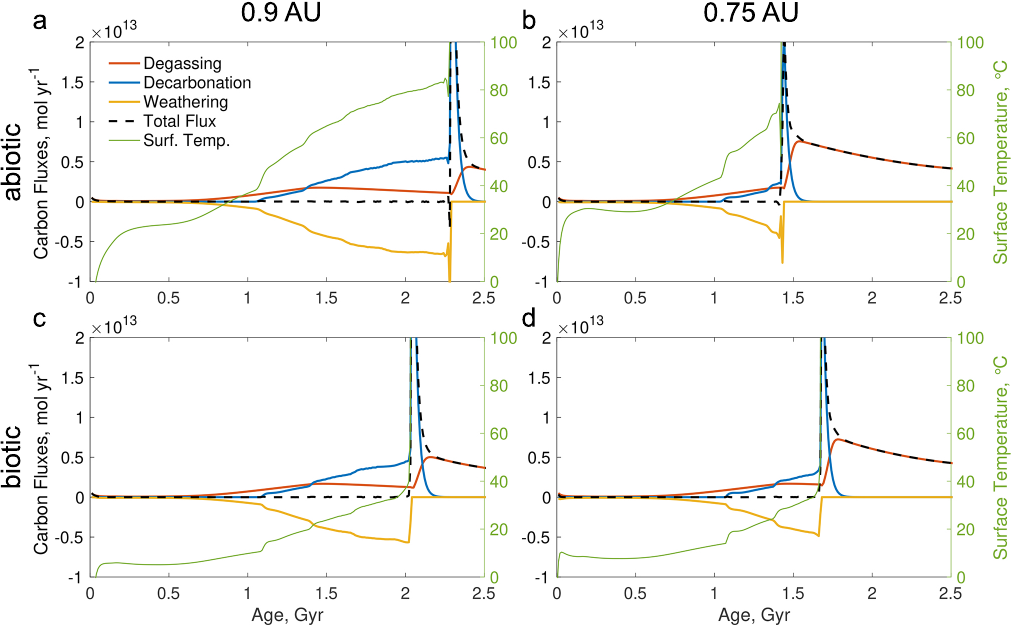}
    \caption
    {Interior-atmosphere carbon fluxes and surface temperature. We used a mantle oxidation state of IW+0 and a stellar distance of d=0.9 AU (left) and 0.75 AU (right) for an abiotic planet (top) and for a biotic planet (bottom). We show mantle carbon degassing (red), crustal decarbonation (blue), weathering (yellow), combined carbon fluxes (black), and surface temperature (green, right axes).}
    \label{fig:fluxes}
\end{figure*}

A main reason for the varying degree of influence of the effect of the biosphere on the habitable timespan is the trade-off between the biologically enhanced rate of weathering (enhancing carbon removal from the atmosphere) and the indirectly enhanced rate of crustal decarbonation (supplying CO$_2$ to the atmosphere). Importantly, this trade-off is controlled by the accumulated mass of carbon that has been stored in the crust throughout the planet's history: the rate of crustal decarbonation depends on the integrated rate of -- biologically enhanced -- weathering. In Fig. \ref{fig:fluxes}, we depict this trade-off between mantle degassing, weathering and decarbonation (top panels: abiotic, bottom panels: biotic). We illustrate the cases for a mantle oxygen fugacity of IW+0 and orbital distances of 0.9 AU (Fig. \ref{fig:fluxes}, left panels) and 0.75 AU (Fig. \ref{fig:fluxes}, right panels), representing two specific cases where the presence of the biosphere reduces (left panels) and extends (right panels) the habitable timespan of the planet.

During the first $\approx$1 Gyr, crustal decarbonation (blue) is zero and mantle carbon degassing (red) is fully compensated by weathering (yellow) for all planets. Subsequently, crustal decarbonation sets in, and the combined atmospheric carbon source fluxes (mantle degassing plus crustal decarbonation) are compensated by weathering. Depending on the stellar distance and on the presence of a biosphere, surface water evaporates between $\approx$1.5 Gyr (panel b) and 2.4 Gyr (panel a). Weathering ceases, CO$_2$ rapidly accumulates in the atmosphere, surface temperature rises, and crustal decarbonation speeds up. The sharp negative peaks of the yellow curves in panels a and b are of numerical origin. After the runaway greenhouse, the rate of mantle degassing increases due to the high surface temperature, which causes thinning of the stagnant lid and shallower melting.

A striking effect of the rate of biologically enhanced weathering is on the surface temperature. While for the abiotic planet (Fig. \ref{fig:fluxes}, top) the surface temperature rises strongly with time and more or less evenly until reaching $\approx80^\circ$C (followed by the onset of the runaway greenhouse), the rise of the surface temperature of the biotic planet (Fig. \ref{fig:fluxes}, bottom) is slowed down by the temperature-dependence of the biological productivity. Therefore, the biotic planet maintains a moderate surface temperature ($\leq\approx35^\circ$C) for almost the entire habitable period, which is followed by a dramatic temperature rise as the bioproductivity becomes less effective at higher temperatures (compare Eq. \ref{eq:biow2}) and ultimately by the transition into the runaway-greenhouse regime.

\subsection{Observational signatures of bioproductivity}
\label{sec:results2}

In section \ref{sec:results1}, we demonstrated that the atmospheric CO$_2$ concentration of a planet near the inner edge of the habitable zone depends on whether the planet is inhabited. Under the condition that other planet-specific parameters could be inferred with sufficient accuracy, this finding presents an intriguing potential observable for future space telescopes.

In the following, we study features in the atmospheric spectrum indicating whether or not the planet underwent a runaway greenhouse in its history, which in turn depends on the presence or absence of a biosphere. We again used our baseline planet (Fig. \ref{fig:co2}) and a stellar distance of 0.9 AU. We illustrate the spectra for the abiotic and biotic planet for two points in time: at 3 Gyr and 4 Gyr. These parameter combinations are well suited for illustration purposes as they cover the transition (both the abiotic and biotic planet are habitable at 3 Gyr, but only the biotic planet is habitable at 4 Gyr).

\begin{table*}
\caption{Scenarios to model the atmospheric spectra.}
\label{tab:2}
\centering
\begin{tabular}{l l l l l l}
\hline
Scenario & Description & Insolation & $P_{CO_2}$ [bar] & $P_{H_2O}$ [bar] & $x_{H_2O}$ \\
\hline
Dry 1 & Biotic, 3 Gyr & 1481.2 W m$^{-2}$ & $1.366\cdot 10^{-1}$ bar & $4.333\cdot 10^{-2}$ bar & $\sim 10^{-6}$ $^{(a)}$\\
Moist 1 & Biotic, 3 Gyr & 1481.2 W m$^{-2}$ & $1.366\cdot 10^{-1}$ bar & $4.333\cdot 10^{-2}$ bar &  $10^{-1}- 10^{-5}$ $^{(a)}$\\
Dry 2 & Biotic, 4 Gyr & 1602.2 W m$^{-2}$ & $1.186\cdot 10^{-1}$ bar & $4.968\cdot 10^{-2}$ bar & $\sim 10^{-6}$ $^{(a)}$\\
Moist 2 & Biotic, 4 Gyr & 1602.2 W m$^{-2}$ & $1.186\cdot 10^{-1}$ bar & $4.968\cdot 10^{-2}$ bar & $10^{-1}- 10^{-5}$ $^{(a)}$\\
Steam 3 & Abiotic, 3 Gyr & 1481.2 W m$^{-2}$ & $2.082\cdot 10^{-1}$ bar & $3.755\cdot 10^{-1}$ bar & 23.7\%\\
Steam 4 & Abiotic, 4 Gyr & 1602.2 W m$^{-2}$ & $8.513$ bar & $2.715\cdot 10^2$ bar & 96.6\%\\
Desiccated 4 & Abiotic, 4 Gyr & 1602.2 W m$^{-2}$ & $8.513$ bar & 0 bar & 0\%\\
\hline
\end{tabular}
\flushleft{Notes: $x_{H_2O}$ is the water volume mixing ratio. (a) \ce{H2O} abundances in the atmosphere are modified by the assumed hydrological cycle. We note that the atmosphere model \texttt{1D TERRA} starts 0.5~km above the modelling domain of the geophysical model (see Section~\ref{sec: PBL}). For all scenarios, we assumed an oxygen fugacity of IW-0.2 and an orbital distance of 0.9 AU from a Sun-like star. We assumed a background gas for all models of 1 bar N$_2$ and for the biotic model runs additionally 0.2 bar O$_2$.}
\end{table*}

We used the atmosphere biochemistry model 1D-TERRA for the biotic scenarios to explore the impact of bioproductivity on atmospheric \ce{CH4} and on the hydrological cycle as \ce{CO2} and thus water vapour increase. In order to cover uncertainties related to modelling the hydrological cycle, we calculated two versions for the biotic scenarios 1\&2 with 1D-TERRA (section~\ref{sec: TERRA}): the dry version reflects an efficient hydrological cycle with the majority of the water vapour already condensing near the surface -- similar to current Earth, whereas the moist version reflects a scenario in which a significant water fraction diffuses into the stratosphere. The moist scenario is thus directly motivated by the outcome of our geophysical model evolution model (see also Appendix \ref{sec: PBL}). Furthermore, for a general overview of the two scenarios, the reader is referred to Table~\ref{tab:2}. For an overview of the model 1D-TERRA \citep{Wunderlich20}, the reader is referred to Appendix~\ref{sec: TERRA}. We note that our moist scenario yields a much higher humidity (by about a factor of 100 in volume mixing ratio) at the near-surface level than the `wet' scenario investigated in \citet{Wunderlich20}.

\begin{figure}
    \centering
 \includegraphics[width=0.48\textwidth]{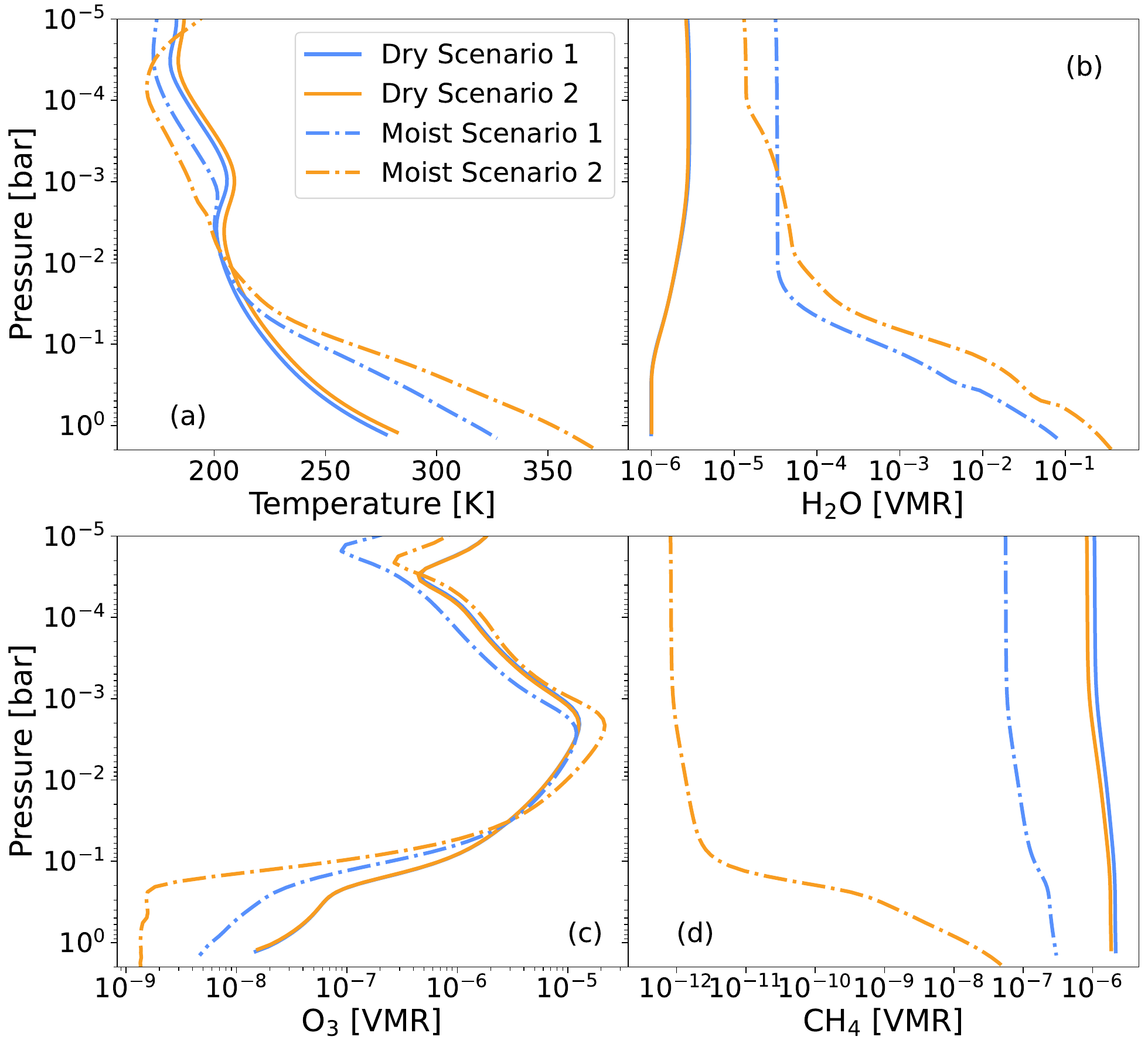}
    \caption{Model results from 1D-TERRA. We show (a) pressure-temperature profiles with efficient near surface condensation and (b--d) abundances or volume mixing ratios (VMR) for (b) \ce{H2O}, (c) \ce{O3}, and (d) \ce{CH4}. The biotic scenarios 1 and 2 are shown in blue and orange, respectively, dry scenarios are denoted by solid lines and moist scenarios by dashed-dotted lines.}
    \label{fig:PT_PTERRA}
\end{figure}

Figure~\ref{fig:PT_PTERRA} shows that the dry scenarios 1 and 2 yield similar results with an ozone driven temperature inversion in the stratosphere. Here, the buffering effect of biological enhanced weathering is very strong, keeping the planet habitable despite higher irradiation. Conversely, the moist scenarios 1 and 2 show higher surface temperatures compared to the dry models due to larger amount of \ce{H2O} vapour in the atmosphere, which is an efficient greenhouse gas. For the moist scenario 2, the surface may become even too hot to remain habitable. In addition, the higher moistness of the stratosphere diminishes the stratosphere temperature inversion (moist scenario 1) or can even remove it entirely (moist scenario 2) such that the stratosphere becomes cooler compared to the dry biotic scenarios.

In the moist scenarios, the abundance of H$_2$O in the atmosphere is larger by at least one order of magnitude compared to the dry scenarios and is particularly high for the lower troposphere ($p>0.1$~bar).  Consequently, \ce{CH4} is diminished by reactions with \ce{OH}-radicals in the moist scenarios near the surface. The effect is strongest for the moist scenario 2, for which very high water volume mixing ratios (VMR) $>0.1$ lead to a very high destruction rate by several orders of magnitude for \ce{CH4}. In comparison, the \ce{CH4} volume mixing ratios produced by the biomass in the moist scenario 1 are only diminished by one order of magnitude compared to the dry, biotic cases that retain relatively high methane abundances of $2\times10^6$ throughout the atmosphere. We note that \ce{CO2} abundances set near the surface via surface weathering are approximately constant throughout the atmosphere. As noted above, our moist scenarios are 100 times more humid than the wet scenarios in \citet{Wunderlich20}, who therefore could not capture the destruction of \ce{CH4}, which we do observe.

Similar to \ce{CH4}, the abundances of ozone are diminished in the lower troposphere ($p<0.1$~bar) with increasing water vapour abundances. However, the majority of \ce{O3} is photochemically created at higher altitudes, where water abundances are relatively low ($<10^{-4}$) even for the moist scenarios and the latter do not strongly affect O$_3$ abundances. The gradual disappearance of the stratospheric temperature inversion with increasing atmosphere moisture is solely due to the latent heat release of water vapour condensing out of the atmosphere, which is stronger than the radiative heating by ozone.

The chemistry results from 1D-TERRA suggest that ozone remains a robust biosignature and that \ce{CH4} abundances can remain high unless water vapour abundances exceed more than 10\% in the lower troposphere. This does, however, not necessarily mean that these two molecules can be detected with JWST on a rocky exoplanet at the inner edge of the host star's habitable zone like TRAPPIST-1e. To explore the observability of the biotic scenarios 1 and 2 with JWST/NIRSpec and JWST/MIRI, we used the abundances and pressure-temperature profiles from \texttt{1D-TERRA} \citep{Wunderlich20} as input parameters for the radiative transfer calculations with \texttt{petitRADTRANS} \citep{Molliere2019} and opacity sources with \ce{N2} pressure broadening ($p_{\ce{N2}}=1$~bar) and added continuum as listed in Table~\ref{Tab:Opacity}. To calculate the noise in the respective JWST observation modes, we employed \texttt{PANDEXO} \citep{batalha2017} and applied the scenarios to TRAPPIST-1e that lies at the inner edge of the habitable zone of its host star. For clarity, we focus on the most promising scenario to detect biosignatures: the dry, biotic scenarios 1 \& 2.

\begin{figure}
    \centering
    \includegraphics[width=0.48\textwidth]{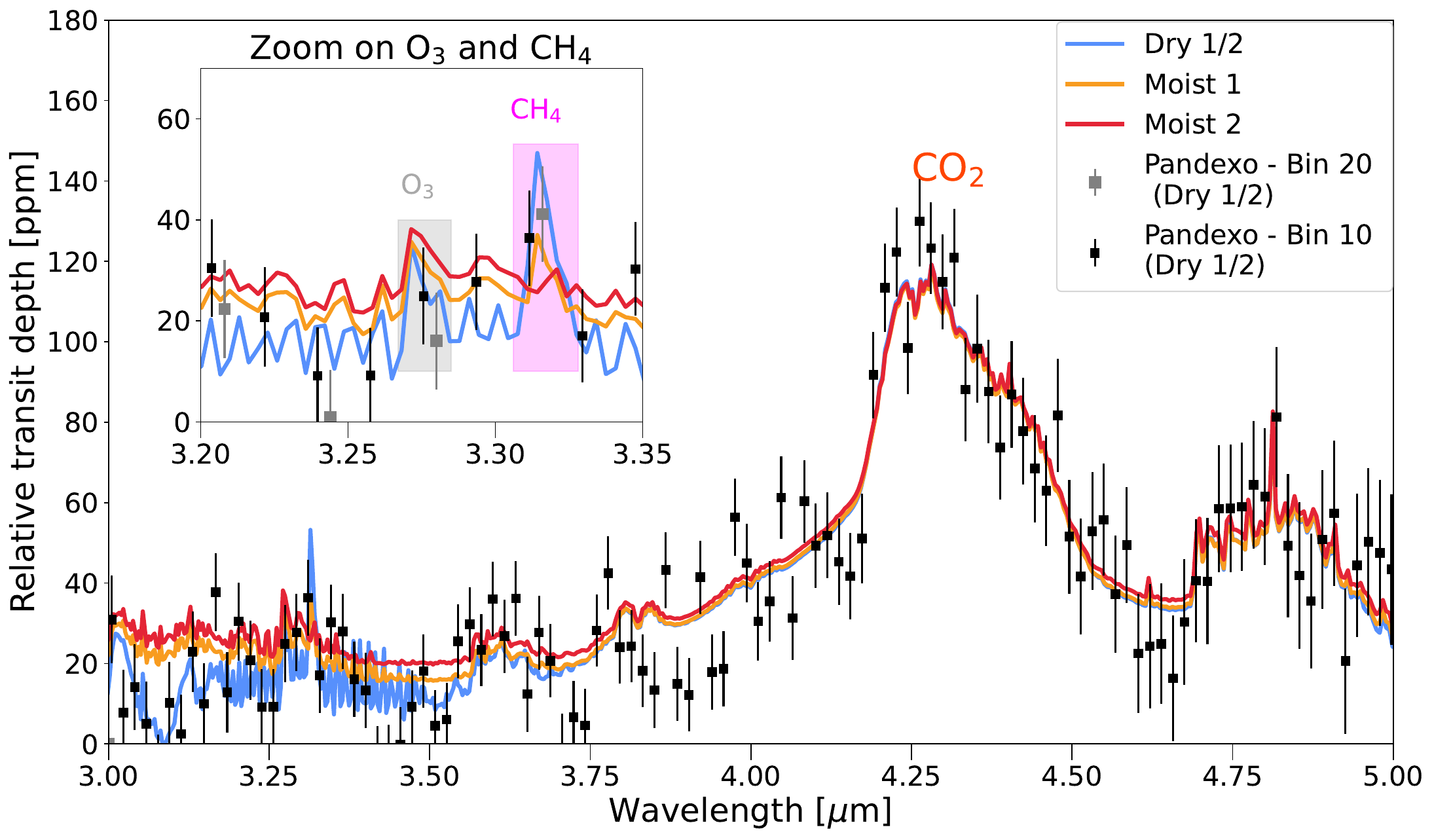}
    \includegraphics[width=0.48\textwidth]{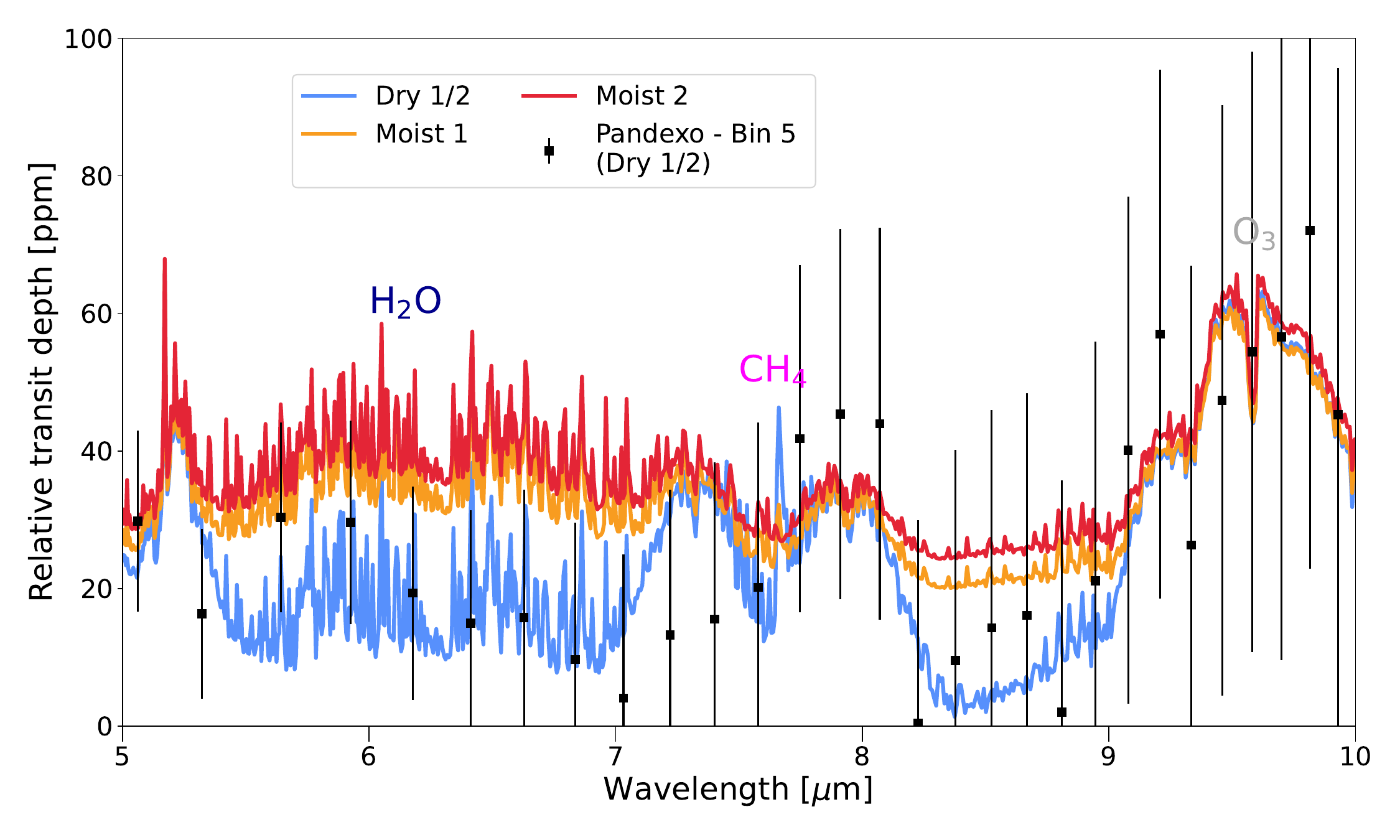}
    \caption{Transmission spectra for the dry and moist biotic scenarios 1 \& 2. We show the JWST/NIRSpec range between 3-5 micron (top) with an inlay for the 3.2 -3.35 micron range to highlight the \ce{O3} (grey box) and \ce{CH4} feature (magenta box). The bottom panel shows transmission spectra for 5-10 micron in the JWST/MIRI range (bottom). We note that the transmission spectra for the dry biotic scenarios 1 and 2 are identical. We further simulated JWST observations with PANDEXO for one scenario (dry 1/2) with 100 transit observations of TRAPPIST-1e (black error bars). We chose the G395M grism setting with 10 pixels per bin (top), 10 and 20 pixels per bins (inlay), and MIRI LRS setting with 5 pixels per bin (bottom panel).}
    \label{fig: PTERRA_Transmission}
\end{figure}

If 100 transit observations are invested, then the strong \ce{CO2} line can be well resolved with JWST/NIRSpec for TRAPPIST-1e, where we find an accuracy of 10~ppm for 100 transits, which is more conservative than the results of \citet{Lustig2022} who derived even better accuracy of 7-8 pm in the 3-5 micron range (Figure~\ref{fig: PTERRA_Transmission}). Conversely, the narrow O$_3$ biosignature is the strongest for the dry scenarios at 3.27~micron in the NIRSpec range and 9.8~micron in the MIRI range. In the NIRSpec range, the \ce{O3} signal is very narrow and lies just left of the stronger \ce{CH4} signature. In the JWST/MIRI range, the \ce{CH4} and \ce{O3} signatures are broader and farther apart. Unfortunately, even with 100 JWST observations, we estimate an accuracy of 20~ppm for JWST/MIRI and thus both the \ce{O3} and \ce{CH4} spectroscopic feature amplitudes are only at $1 \sigma$ in the cloud-free case. Investing 100 transit observations using MIRI with its lower data accuracy does therefore not appear to be promising to identify the signatures of the biotic atmosphere scenarios.

A close-up of the \ce{O3}--\ce{CH4} double feature in the NIRSpec range (Figure~\ref{fig: PTERRA_Transmission}, inlay) clearly shows that \ce{O3} has a signal strength between 1--2 $\sigma$ for all scenarios. The higher bioproductivity of scenarios 1 and 2 (see Appendix~\ref{sec: TERRA}) lead, however, to a relatively high \ce{CH4} signature of 40~ppm for the dry scenarios that is well above 3$\sigma$ with 100 JWST observations when neglecting the impact of clouds. However, achieving an accuracy of 10~ppm with the observational data requires reducing the wavelength resolution down to 10 -- 20 pixels per bin, which resolves the \ce{CH4} signal with only 1 -- 2 points. With increasing water vapour content in the atmosphere, however, \ce{CH4} will be destroyed near the surface and the signal drops to $1\sigma$ already for the moist scenario 1. For even more water vapour, \ce{CH4} is not observable.

\begin{figure}
    \centering
    \includegraphics[width=0.48\textwidth]{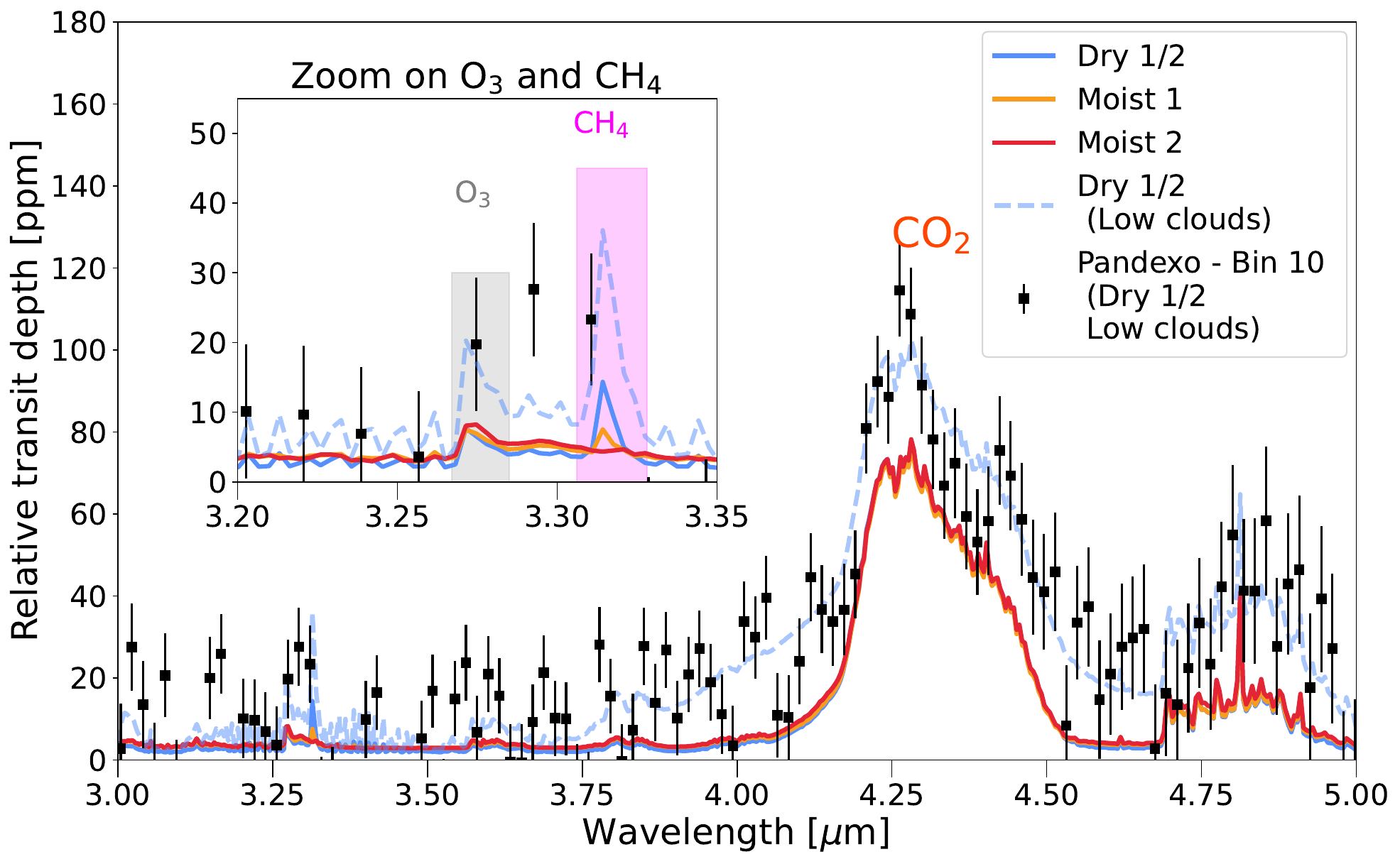}
    \caption{Transmission spectra for the biotic scenarios. We show the JWST/NIRSpec range between 3 -- 5 micron with a continuous grey cloud deck at the $p=0.01$~bar (solid lines) and the dry biotic scenarios 1 \& 2 with a lower cloud deck at the $p=0.1$~bar (dashed line). The inlay for the 3.2 -- 3.35 micron range highlights the \ce{O3} (grey box) and \ce{CH4} feature (magenta box). We further simulated JWST observations with PANDEXO for the dry biotic scenarios 1 \& 2 with the lower cloud deck, simulating 100 transit observations of TRAPPIST-1e (black error bars). We chose the G395M grism setting and combined 10 pixels in one bin.}
    \label{fig: PTERRA_Transmission_Clouds}
\end{figure}

We further tested the impact of clouds on the biosignatures for the NIRSpec transmission spectra, which appears to be most promising to identify a habitable, biotic planet with JWST. Such observations will support statistical studies of the extension of the habitable zone \citep{lehmer2020carbonate,schlecker2023}. We first assumed a grey cloud deck at $p=10^{-2}$~bar. This scenario is roughly equivalent to a thick tropical cumulonimbus thundercloud and therefore represents the worst case scenario for a dry Earth-like atmosphere. Figure~\ref{fig: PTERRA_Transmission_Clouds} shows that the \ce{CH4} signal strength that was found to be above $3\sigma$ in the cloud-free, dry scenario for 100 JWST observations drops down to 10~ppm or 1~$\sigma$ with thick thundercloud coverage. The \ce{O3} signature at 3.3 micron remains visible because the majority of the ozone layer is well above the cloud top (Figure~\ref{fig:PT_PTERRA}). The amplitude of the ozone line is, however, diminished by a factor of two and is thus below $1\sigma$ with 100 transit observations for a cloudy atmosphere.

We also tested a more benign case with a cloud top at 0.1~bar for the dry biotic scenarios, which would correspond to a thinner stratocumulus cloud for an Earth-like atmosphere. Here, the \ce{CH4} signature drops from 40~ppm to 30~ppm, which is still at $3\sigma$ even with our conservative noise estimate for 100 JWST observations. Again, the necessary reduction in wavelength resolution to achieve 10~ppm accuracy leaves at most 2 points to resolve the \ce{CH4} signature.

For all cloud scenarios, the amplitude of the \ce{CO2} feature remains well above $3~\sigma$ even in this thick cloud case and should thus allow to constrain the state of the silicate weathering cycle as outlined by \citet{Lustig2022}. The 4.25~micron \ce{CO2} feature ($P_{CO_2}=(1.196 -1.366)\cdot 10^{-1}$~bar) appears to be very robust against changes in water vapour abundances and pressure--temperature profile variations of up to 100~K at the surface. It is clearly identifiable with 100 transit measurements even with a relatively high cloud deck. The 4.25~micron \ce{CO2} feature can thus be seen as a reliable indicator of \ce{CO2} abundances \citep[see also][]{Lustig2022}.

Given the large time investment that is needed to characterise atmospheres of rocky planets, we also explored signatures of other atmosphere scenarios that signal the end of habitability. To this aim, we constructed two scenarios with surface temperatures that are too high to sustain habitability and that would also result in total failure of the hydrological cycle (steam-dominated atmosphere scenarios 3\&4). A steam-dominated atmosphere is, however, expected to result in a desiccated \ce{CO2}-dominated atmosphere within 100~million years (see also Table~\ref{tab:2}). These scenarios are explored as supplemental information in the Appendix (\ref{sec: simple atmo}, \ref{sec: observe_end} and \ref{sec: Observe_end_cloud}).

\section{Discussion}
\label{sec:discussion}

How does biological enhancement of weathering affect the habitable timespan of exoplanets? Could we use this insight to advance the search for life beyond Earth? To answer these questions, we modelled the coupled interior-atmosphere evolution of stagnant-lid planets accounting for biologically enhanced weathering and compared the evolution of the atmospheric CO$_2$ partial pressure between a biotic and an abiotic planet for different planet compositions and orbital distances. Following this, we compared the resulting emission and transmission spectra between these planets.

\subsection{Impact of bioproductivity and other planet-specific parameters on atmospheric CO$_2$}

Our results indicate that the biosphere has the potential to significantly extend the habitable period. For planets located near the inner edge of the habitable zone, with similar planet parameters (such as interior composition), the CO$_2$ signature of inhabited planets with active weathering would be distinguishable from that of abiotic planets that have undergone a runaway-greenhouse effect. However, it is important to note that our results are based on forward modelling with specific, well-defined input parameters. In particular, mantle oxygen fugacity, initial mantle temperature, and reference mantle viscosity play crucial roles. If the atmospheric CO$_2$ signature of an exoplanet is observed without information about these parameters, it is not possible to directly link the CO$_2$ signature to the presence or absence of a biosphere. Improved constraints on these parameters will help resolve this ambiguity, which we address in the following.

Inferring the mantle oxygen fugacity of exoplanets with sufficient accuracy is a primary challenge in isolating the effect of life on the habitable timespan. While a biosphere for a moderately reduced mantle (IW-0.4) substantially extends the habitable timespan across all explored stellar distances, for a more oxidised mantle (IW+0), the biosphere extends the habitable timespan only for planets close to their host star and only to a limited extent. This suggests that planets with high incident insolation are the most suited to a use biologically prolonged habitable period as a biosignature.

One way to infer a planet's oxidation state is by analysing the observed ratio between reduced species (\ce{CO}, \ce{H2}) and oxidised species (\ce{CO2}, \ce{H2O}) \citep[e.g.][]{ortenzi2020,liggins2022}. Another complementary approach is to study the planet's host star, as both the planet and the star are formed from the same primordial material \citep[e.g.][]{adibekyan2021, guimond2023}. However, as pointed out by \cite{guimond2023}, even with the same \ce{Fe3+}/$\sum$Fe ratio, the mantle oxidation state can vary significantly depending on its bulk composition. Finally, ongoing advancements in exoplanet atmosphere characterisation will further enhance our understanding of the diversity of mantle oxidation states \citep{ortenzi2020}.

In addition to the mantle oxidation state, the initial mantle temperature and the reference mantle viscosity are important abiotic factors that influence the habitable timespan of a stagnant-lid planet by controlling the early outgassing rate of CO$_2$. Planetary accretion models can provide insights into the initial thermal energy, and thus the initial mantle temperature, for planets in general \citep[e.g.][]{Schubert:2001}. In contrast, the reference mantle viscosity could vary between planets in particular due to differences in the volatile content in the mantle \citep[e.g.][]{karato1986}. Models of planetary system formation can offer insights into the bulk volatile concentration of exoplanets \citep[e.g. ][]{kamp2013uncertainties,moriarty2014chemistry}, making them highly relevant in this context.

Parameter uncertainties explored in this study highlight the need for future research, particularly focused on refining our understanding of exoplanet composition and initial mantle temperature, to effectively use the biologically extended habitable period as a meaningful biomarker. Nevertheless, even if uncertainties related to abiotic factors remain difficult to reduce and may overshadow biosphere-related effects, the framework presented here provides a valuable approach for selecting the most promising exoplanets for follow-up studies: in particular, planets near the inner edge of the habitable zone with low atmospheric CO$_2$ levels are strong candidates to have an active biosphere. If planet-specific parameters can be inferred, exoplanets that meet these criteria and also have a reduced mantle ($\leq$ IW-0.2) and high mantle viscosity ($\eta_{ref}\geq$ 10$^{21}$ Pa s) should be prioritised for future long-term observations, as the influence of a biosphere in prolonging the habitable period is especially pronounced under these conditions. While we found that the initial mantle temperature can shift the habitable timespan, its influence on the impact of a biosphere is minimal.

From the biological point of view, a main simplification of our model was that even though we calculated the bioproductivity as a function of temperature and atmospheric CO$_2$ concentration, we neglected any temporal biological evolution. On Earth, land plants, which substantially enhance weathering, have emerged only in the Paleozoic \citep{berner1997rise}. If this also applies to the stagnant-lid planets discussed in this paper, the early period (until the emergence of land plants) of the biotic planet would resemble that of the abiotic planet and so would the crustal carbon concentration. If land plants emerged later, shortly before the planet would otherwise become too hot, biologically-enhanced weathering would efficiently extend the habitable timespan since the decarbonation rate would follow the weathering rate with a delay: it takes time until carbonated crust reaches the decarbonation depth, at least about 1 Gyr, and even more in the later evolution, due to the cooler mantle and hence the smaller rate of crustal production \cite[compare][]{honing2019carbon}. We also note that the assumed factor of biological enhancement of weathering on present-day Earth is a conservative estimate; an abiotic Earth would likely have an even smaller reference weathering rate \citep{lenton:2018}.

\subsection{Observability of the signatures of bioproductivity}

Several studies have explored the observability of \ce{CO2} and biosignatures like \ce{CH4} and \ce{O3} under conditions of increased irradiation compared to present-day Earth \citep{Schwieterman2019,Wunderlich20,Lustig2022}. These studies, however, relied on assumptions on methane production based on present-day Earth rates, treated it as a free parameter, and/or did not consider the interplay between the biosphere and atmospheric \ce{CO2}. \citet{Wunderlich20} investigated the impact of increased humidity on \ce{CH4} and \ce{O3} due to changes in the water cycle, but without considering feedback with the biosphere. Table~\ref{tab:CH4-O2} summarises methane, oxygen fluxes and \ce{CO2} from studies most similar to this work. In our study, we derived the biotic \ce{CH4} production rate from the coupled model. Furthermore, we made two extreme humidity assumptions for the atmosphere above the atmospheric planetary boundary layer, which can be seen as the interface region between the geophysical and the atmospheric model employed here (see sections~\ref{sec: TERRA} and \ref{sec: PBL}). We found that the humidity has a strong impact on the atmospheric \ce{CH4} levels and a less pronounced impact on atmospheric \ce{O3}. We also assessed the impact of clouds, identifying scenarios where cloud cover could diminish the observability of \ce{CH4}.

\begin{table}
\caption{Overview of biosignature production.}
\label{tab:CH4-O2}
\centering
\begin{tabular}{l| l l}
  & \ce{CH4} flux [cm${}^{-2}$/s] & \ce{CO2} [bar] \\
\hline
Scenario 1 & $1.25 \cdot 10^{11}$ & $1.366 \cdot 10^{-1}$ \\
Scenario 2 &  $1.14 \cdot 10^{11}$ & $1.186 \cdot 10^{-1}$\\
W2020 & $6.31 \cdot 10^{10}$, $1.12 \cdot 10^{8}$ & $10^{-1}$\\
S2019 & $10^{6} - 10^{14}$ & $4 \cdot 10^{-2}$\\
\end{tabular}
\flushleft{Notes: W2020: \citet{Wunderlich20}; S2019: \citet{Schwieterman2019}. The \ce{O2} flux is $1.21 \cdot 10^{12}$ cm$^{-2}$/s in all cases. The \ce{CH4} flux used by \citet{Wunderlich20} is the biotically generated flux for present-day Earth without anthropogenic contributions.}
\end{table}

Our results show that the \ce{CO2} absorption band at 4.3 micron is well resolvable with 100 transit observations with JWST/NIRSpec. This is in accordance with \citet{Lustig2022}, who studied the potential observability of habitability of TRAPPIST-1e. Between 3.2 and 3.4~micron, \ce{O3} and \ce{CH4} features, in principle, allow to identify the habitable, biotic scenarios 1\&2. However, only the \ce{CH4} signature arising from the increased bioactivity reaches $\approx 3 \sigma$ significance. This holds even true in the presence of a low cloud deck at 0.1~bar. Reaching the necessary 10~ppm accuracy to identify \ce{CH4} requires, however, data binning down to 1--2 points in the narrow \ce{CH4} (3.30~$\mu$m and 3.35~$\mu$m) absorption feature. Thus, a robust \ce{CH4} detection will require careful treatment of systematic and stellar noise.

We stress that we chose conservatively that the noise accuracy is not higher than 10~ppm between 3--4 $\mu$m. \citet{Lustig2022}, on the other hand, simulate an even higher accuracy (8~ppm) with 100~transit measurements in this critical wavelength range, which would raise the amplitude of the \ce{CH4} signature to more than 4 $\sigma$. We further point out a similarly challenging and successful detection in exoplanetary atmospheres: \ce{He} was identified with HST/WFC3 observations with improved data and noise treatment based on a single, binned data point \citep{Spake2018}. Thus, we conclude that while identifying \ce{CH4} with JWST/NIRSpec transit observations might be challenging, it is a worthwhile endeavour if we aim to identify habitable temperate worlds in the near future. With the prolonged habitability due to feedback with the biosphere, we also highlight that the period where \ce{CH4} and \ce{O3} is present in the atmosphere can be significantly extended.

\citet{Ostberg2023} also conclude that the 3.3~$\mu$m \ce{CH4} atmospheric feature is ideally suited to identify a habitable Exo-Earth. The significance of the \ce{CH4} biosignature drops, however, below 1~$\sigma$ in the presence of a thick continuous cloud deck at $p=0.01$~bar. Further, if the water vapour content increases to 1\%-10\% at surface level, then \ce{OH} radicals destroy \ce{CH4}\footnote{The importance of the OH production rate and the photochemical time scales for \ce{CH4} is also pointed out by \citet{Wunderlich20,Grenfell2013,Schwieterman2019}.}. Thus, the observability of biosignatures depends critically on an efficient hydrological cycle and cloud formation as the planet evolves out of habitability. 

In our dry scenarios, the resulting \ce{CH4} and \ce{O3} amplitudes of 60~ppm and 40~ppm are roughly in agreement with the outcome of the most similar scenario with 0.1~bar \ce{CO2} from \citet{Wunderlich20}, though that study did not account for instrumental noise and cloud effects. We found that, without efficient rainout, the geophysical evolution leads to higher humidity than in the wet model from \citet{Wunderlich20}. As a consequence, \ce{OH} radical production becomes efficient enough to destroy \ce{CH4}, even in a low UV environment around M dwarfs, which generally favour methane preservation \citep{Wunderlich2019,Schwieterman2024}. This effect arises once the water vapour volume mixing ratio at the bottom of the atmosphere model exceeds 0.1, which is more than 100 times larger than in the wettest conditions in \citet{Wunderlich20}. Additionally, we observe a sharp shift from \ce{CH4}-preservation to destruction in response to high \ce{H2O} levels, with atmospheric \ce{CH4} abundances dropping by 6 orders of magnitude between moist scenarios 1 and scenario 2 as the water vapour mixing ratio increases from 0.1 -- 0.2 at the bottom (Figure~\ref{fig:PT_PTERRA}). Therefore, our work highlights that even around M~dwarfs, \ce{CH4} might be produced but could become unobservable in extremely humid atmospheres. In contrast, \ce{O3} generated in the upper atmosphere remains relatively unaffected by increased atmospheric humidity. Therefore, constraining atmospheric water abundances along with the detection of \ce{O3} or \ce{O2} would be required to identify a moist biotic scenario.

The inferred \ce{O3} abundances for the biotic scenarios 1 \& 2 with \texttt{1D TERRA} can reach $\approx 10^{-5}$ in the stratosphere, which is similar to Earth's current ozone concentration and also agrees with a recent 3D ozone chemistry model for tidally locked exo-Earths \citep{Braam2022}. For exo-Earths around M dwarf stars, however, a range of possible \ce{O3} scenarios is discussed. On these tidally locked planets, \ce{O3} distribution is highly affected by 3D circulation \citep[e.g.][]{Carone2018,Chen2019,Chen2021,Braam2022, Braam2023}. Consequently, results for \ce{O3} concentrations on Earth-like atmospheres around M dwarfs range between abundances smaller than one order of magnitude \citep{Yates2020} or larger by two orders of magnitude compared to Earth with flare driven ozone production \citep{Chen2021}. The \ce{O3} abundances found in this work would yield only a 2~$\sigma$ \ce{O3} signal at 3.2 micron with 100 JWST transit observations for TRAPPIST-1e, even in the cloud-free case. An enrichment by a factor of 100 in abundance may push the \ce{O3} signal to detectable levels \citep{Barstow2016}. However, the \ce{O3} absorption feature in the 3.27~$\mu$m is even narrower than the \ce{CH4} feature (Fig.~\ref{fig: PTERRA_Transmission}, inlay) and thus any attempt to detect \ce{O3} with JWST/NIRSpec will similarly require highly accurate noise reduction.

Although \ce{O3} may be difficult to observe with JWST, it is, in principle, a relatively stable indicator of biotic atmospheres in our model framework. It is photochemically produced above the cloud deck and its abundances are hardly affected by higher water abundances even in the case of a less efficient water cycle (moist biotic scenarios 1 \& 2). In the case of a cloud-free habitable dayside or a dayside that is only partly covered by clouds, already 10 eclipse observations with JWST/MIRI could yield some insights. Here, the \ce{CH4} biosignature at 7.7 micron (2~$\sigma$) and the \ce{O3} signature at 9.5 micron (3~$\sigma$) do not overlap. However, in case of a continuous cloud deck at $p=10^{-2}$~bar, the almost isothermal upper atmosphere will make the observation of molecular features impossible.

\citet{Snellen2013} point out that the identification of \ce{O3} and \ce{O2} on potentially habitable worlds could be challenging even with telescopes like the Extremely Large Telescope (ELT) due to telluric contamination. The identification of molecular \ce{O2} in the optical, specifically in the 0.76~$\mu$m band, could be more promising. In this band, telluric oxygen could be differentiated from oxygen in the atmosphere of the observed planet. However, \ce{O2} alone without the identification of \ce{H2O}, \ce{CO2} and ideally also \ce{CH4} is not a definitive biomarker for rocky planets around M dwarf stars, where efficient XUV photolysis of \ce{H2O} can easily create \ce{O2} abiotically \citep{Luger2015, Meadows2018}. High resolution spectroscopy ($R \sim$ 100.000) observations of \ce{O2} with ground based telescopes like ELT may help, however, to distinguish between biotic and abiotic scenarios for rocky planets, for which \ce{H2O} and \ce{CO2} has been found with space based telescopes. Spectrographs on space telescopes are typically limited in spectral resolution ($R < $10.000) and thus not well suited to uniquely identifying the relative narrow \ce{O2} lines \citep[e.g.][]{Lin2022}. \cite{Lin2022}, specifically, point out that the ELT combined with the JWST can be highly useful in assessing of habitability on TRAPPIST-1e.

Cloud coverage depends on the 3D climate state of the rocky planets. Tidally locked 3D exo-Earth scenarios for TRAPPIST-1e consistently predict a climate state with a strong upwelling branch over the dayside \citep[e.g.][]{Carone2018,Braam2022,krissansen2022understanding}, where eclipse observations would probe. The resulting thick dayside clouds would render observations difficult and eclipse data would be consistent with a feature-less black body curve between 200~K and 320~K, depending on the upper atmosphere temperatures. The 200~K scenarios would be valid for the biotic, habitable scenarios but also for a desiccated \ce{CO2} atmosphere after the planet has been rendered uninhabitable. The 320~K upper atmosphere scenarios are steam--dominated atmospheres or runaway-greenhouse atmosphere scenarios \citep[e.g.][]{goldblatt2013,schlecker2023,Barth21,boukrouche2021}.

Clouds complicate the observability of rocky planets \citep[e.g.][]{Komacek2020,Ostberg2023,Cohen2024}. We note, however, that the coverage of high-altitude hazes in biotic Earth-like atmospheres is not expected, because for such a scenario \ce{CH4}/\ce{CO2} ratios greater than 1 would be required \citep{Arney2016}. For the desiccated scenario 4, we did not explore the impact of hazes further because the grey cloud deck scenario already renders such an atmosphere difficult to constrain. \ce{SO2} as a signature of geophysical activity might be promising for exo-Venus identification \citep{Ostberg2023}.

Three-dimensional exo-Earth climate simulations for TRAPPIST-1e predict that the terminator regions, where transmission spectra probe, should be virtually cloud-free \citep{Sergeev2022Bistatic}. Thus, 100~JWST transit observations with NIRSpec are a worthwhile time investment to aim at constraining at least \ce{CO2} and \ce{CH4}. This is true, in particular, since the 4.3~micron \ce{CO2} absorption band remains a robust feature of oxidised exoplanet atmospheres also beyond habitable scenarios. If paired with constraints of \ce{H2O} signatures at shorter wavelengths ($1-3$~micron), JWST transmission observations may also help constrain the humidity of the planet's climate. This will require, however, an investment in minimising the impact of the strong variability of the host star \citep{Rackham2023}.

Despite all the observational challenges, observations of planets near the inner edge of the habitable zone like TRAPPIST-1e will pave the way to identify potentially biotic planets with a suitable range of \ce{CO2} abundances, to re-assess the limits of the habitable zone, and therefore to maximise our chances to characterise a habitable and potentially inhabited world in the near future. This requires, however, also to recognise moist biotic scenarios as outlined in this work.

We further point out that non-tidally locked rocky exoplanets, on which the space mission \texttt{PLATO} will focus, would potentially exhibit a non-continuous cloud coverage. The proposed \texttt{LIFE} mission is ideally suited to characterise these non-tidally locked planets in the habitable zone of FGK stars \citep{Alei2022}.

Last but not least, we stress that rocky exoplanet atmospheres are complex and that various effects may lead to similar observational signatures. The case of the temperate sub-Neptune K2-18b serves as an example that the atmospheres of temperate worlds can in principle be characterised and also acts as a cautionary tale. The recently obtained JWST observations of \ce{CH4} and \ce{CO2} in a reducing \ce{H2}-dominated atmosphere with a notable absence of \ce{H2O} call for disequilibrium chemistry models that connect deeper parts of the planet with the observable atmosphere \citep{Maddu2023}. However, a magma ocean below the atmosphere and a liquid water layer both can explain the JWST observations \citep{Shorttle2024}. Similar results could also arise for temperate rocky planets. Thus, it remains important to determine the potential complexity of these planets with a diverse set of models. One important factor is the impact of a biosphere and potential observable signatures as discussed in this work.

\section{Conclusions}
\label{sec:conclusions}

This study is the first to directly link the atmospheric CO$_2$ signature of planets near the inner edge of the habitable zone to the presence or absence of a biosphere. Specifically, we find that planets with low atmospheric CO$_2$ in this region are more likely to be inhabited, as biological processes efficiently regulate CO$_2$ levels through enhanced weathering and thereby postpone the runaway greenhouse and the accompanying dramatic rise of atmospheric CO$_2$. To reach this conclusion, we modelled the coupled interior-atmosphere evolution of stagnant-lid planets, accounting for biologically enhanced weathering, carbonate burial, and decarbonation. The resulting abundances of atmospheric gases were then used to simulate the spectral signatures.

For most parameters explored in this study, the presence of a biosphere significantly prolongs the habitable period of a planet, by up to approximately two billion years. The transition to an uninhabitable state is marked by an increase in atmospheric CO$_2$ partial pressure by roughly two orders of magnitude. As a result, planets with similar geophysical characteristics and orbital distances would exhibit vastly different CO$_2$ levels depending on whether they are inhabited or not.

However, biological enhancement of weathering is not the only factor controlling the habitable timespan. Mantle oxygen fugacity, initial mantle temperature, and reference mantle viscosity also play crucial roles. Isolating the effect of a biosphere on the habitable timespan to use it as a biomarker remains challenging, as it requires narrowing down these planet-specific parameters for observed targets. Nevertheless, even if abiotic factors cannot be completely constrained and may obscure the biosphere's influence, the framework presented here still provides valuable insights for identifying the most promising candidates to have an active biosphere. These candidates could then be prioritised for future long-term space-telescope observations. In particular, planets near the inner edge of the habitable zone with low atmospheric CO$_2$ would be of first-order interest. Additionally, planets with a reduced mantle and a high mantle viscosity are promising, as these conditions amplify the effect of a biosphere in extending the habitable period. Under these circumstances, low levels of atmospheric CO$_2$ are even more likely to indicate the presence of life.

From an observational perspective, \ce{CO2} serves as a key signature of active surface weathering and dominates the atmospheric spectrum at 4.3~microns. This feature should be readily detectable with 100~JWST transmission measurements, even under cloudy conditions, provided that the atmosphere is oxidised enough to suppress high-altitude hazes \citep{Arney2016}. Additionally, increased bioproductivity could yield a detectable \ce{CH4} biosignature at 3.3 micron. However, the observability of CH$_4$ may be hindered by an inefficient hydrological cycle that enriches the atmosphere with water vapour, as \ce{OH} radicals can destroy \ce{CH4}. We also find that the destruction of \ce{CH4} is triggered when the water vapour volume mixing ratio exceeds 0.1 near the surface. At lower values, \ce{CH4} abundances remain at least at 0.5 ppm throughout the atmosphere and reach 2 ppm for the dry scenarios and increased bioactivity. \ce{O3} levels are more stable, even in our moist biotic scenarios, and cloud presence does not significantly hinder the detectability of \ce{O3}. Detecting the latter is likely challenging with 100 transmission observations, but already ten  eclipse observations could constrain both \ce{CH4} and \ce{O3}, assuming the planet's dayside is partially cloud-free.

Tidally locked rocky exo-Earths are expected to have continuous dayside cloud coverage. Therefore, additional ground-based observations with high-resolution telescopes such as the ELT could complement space-based data, helping to identify biosignatures before the launch of the Habitable World Observatory (HWO). In the immediate future, resolving the 4.3 micron \ce{CO2} feature and potentially \ce{H2O} using JWST transmission spectra of potentially habitable planets such as TRAPPIST-1e offers a promising pathway for atmosphere characterisation \citep[see also][]{Lustig2022}.

In summary, the control of bioproductivity on the onset of the runaway greenhouse near the inner edge of the habitable zone in combination with the observational \ce{CO2} signatures accompanying the transition from a habitable to an uninhabitable world, as well as the determination of atmospheric \ce{CH4} and \ce{O3} in conjunction with a monitoring of water abundances, are promising pieces of the puzzle aiming to assess whether or not an exoplanet is inhabited. Future work is needed both to improve our understanding of the geological and geochemical state of exoplanets and to advance the development of the next generation of space telescopes.

\begin{acknowledgements}
We thank Brad Foley for valuable comments on a previous version of this manuscript. L.C. acknowledges support by the DFG priority programme SP1833 `Building a habitable Earth' Grant CA 1795/3, the Royal Astronomical Society University Fellowship URF R1 211718 hosted by the University of St Andrews, and the European Union H2020-MSCA-ITN-2019 1136 under Grant Agreement no. 860470 (CHAMELEON). J.L.G and N.I thank the German Research Foundation (DFG) for financial support via the project The Influence of Cosmic Rays on Exoplanetary Atmospheric Biosignatures (Project number 282759267). KH acknowledges the FED-tWIN research program STELLA (Prf-2021-022), funded by the Belgian Science Policy Office (BELSPO). P.B. and N.T. acknowledge support from the DFG through the priority program SPP1992 `Exploring the Diversity of Extrasolar Planets', grant TO 704/3-1.
\end{acknowledgements}
\bibliographystyle{aa}
\bibliography{literature.bib}
\begin{appendix}

\section{Testing the observability of the end of habitability with a simplified atmosphere model}
\label{sec: simple atmo}

\begin{figure}
    \centering
    \includegraphics[width=0.48\textwidth]{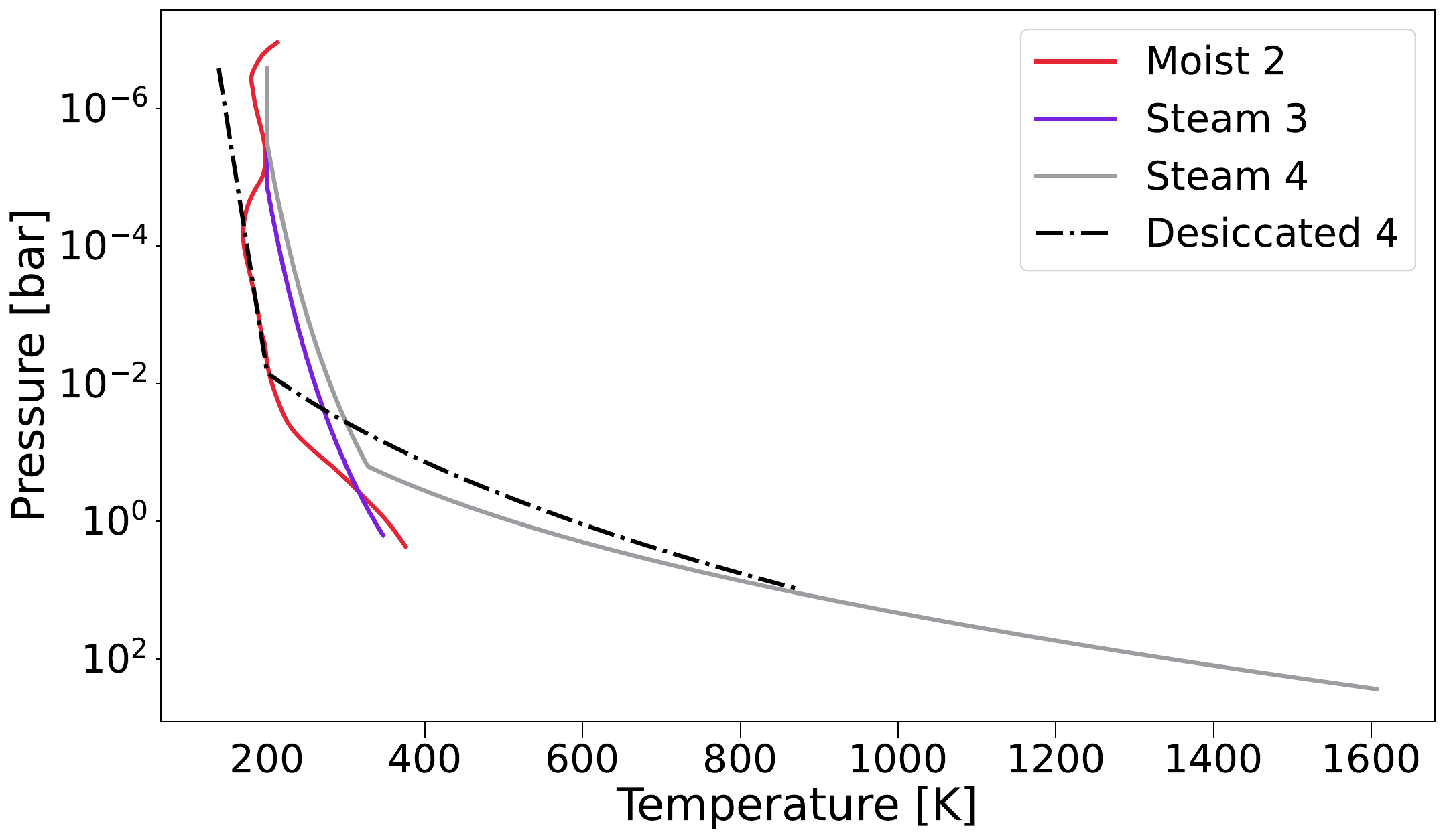}
    \caption
    {Pressure--temperature profiles used to explore the end of habitability. We show steam-dominated scenarios 3 (purple) and 4 (grey) and a desiccated scenario 4 (black dashed-dotted). For comparison, the moist scenario 2 is shown in red.}
   \label{fig:MoistGH_Temp}
\end{figure}

In this section, we use a simple atmospheric model to characterise a rocky planet evolving out of habitability. Figure~\ref{fig:MoistGH_Temp} shows the two-step pressure--temperature profiles for the abiotic scenarios 3 and 4 (Table~\ref{tab:2}), where we followed the approach of \citet{graham2020} to calculate the pressure-temperature profile. We assumed that the entire surface water reservoir has evaporated and that the planet enters a steam-dominated state with vigorous vertical mixing. For such a state, we further assumed equilibrium between \ce{H2O} evaporation and condensation as well as uniform \ce{H2O} volume mixing ratios throughout the atmosphere. This results in water volume mixing ratios of 24\% in scenario 3 and 97\% in moist scenario 4. Consequently, the pressure-temperature profile follows mostly a moist adiabat in the troposphere above the surface. For the very extended atmosphere in moist scenario 4, the pressure-temperature profile starts with a dry adiabat and enters a moist adiabat, dominated by water condensation for pressures $\leq 1$~bar (Figure~\ref{fig:MoistGH_Temp}).

\begin{table}[b]
        \centering
        \caption{Details on opacity sources.}
        \begin{tabular}{p{5cm} l l} 
        \hline
        Data type & Data source &  Ref. \\
        \hline
        H$_2$O broadened by 1 bar N$_2$ (biotic$^c$), H$_2$O and CO$_2$$^a$ (abiotic) & HITRAN2020 & $^{(1)}$ \\
    CO$_2$ broadened by 1 bar N$_2$ (biotic$^c$), CO$_2$ and H$_2$O (abiotic) & HITRAN2020 & $^{(1)}$ \\
\rule{0pt}{3ex} \multirow{ 4}{*}{H$_2$O continuum} & CAVIAR$^b$  & $^{(2,3,4)}$\\
 & MT\_CKD & $^{(5)}$ \\
& Baranov 2008 &  $^{(6)}$ \\
& Odintsova 2020 & $^{(7)}$ \\
\rule{0pt}{3ex}CO$_2$ continuum & MT\_CKD & $^{(8)}$ \\
        O$_3$ broadened by 1 bar N$_2$ & HITRAN2020 &  $^{(1)}$ \\
    CH$_4$ broadened by 1 bar N$_2$ & HITRAN2020 &  $^{(1)}$ \\
\hline
        \end{tabular}
        \label{t:opacities}
        \flushleft{Notes: $a$: We used air broadening coefficients as a proxy for CO$_2$ broadening since CO$_2$ broadening of H$_2$O parameters were not available in HITRAN2020; $b$: Details on the H$_2$O continuum data used from the CAVIAR laboratory experiment can be found in \cite{22AnChEl}; $c$: Biotic cases contain 0.2~bar \ce{O2}; $^{(1)}$\cite{22GoRoHa}, $^{(2)}$\cite{11PtMcSh}, $^{(3)}$\cite{16ShCaMo}, $^{(4)}$\cite{09PaPtSh}, $^{(5)}$\cite{12MlPaMo}, $^{(6)}$\cite{08BaLaMa}, $^{(7)}$\cite{20OdTrSi}, $^{(8)}$\cite{12MlPaMo}}
  \label{Tab:Opacity}
\end{table}
 
Steam-dominated atmospheres are subject to strong atmospheric erosion, where an Earth-like planet around a Sun-like star is expected to lose its total water reservoir within 100~million~years \citep[][]{Abe2011,Hamano2013}. For rocky planets in the habitable zone of the active M dwarf TRAPPIST-1, an Earth-like surface water reservoir is expected to be removed even faster \citep[e.g.][]{Barth21,Lincowski2023}. Thus, as an endstate of the planet's evolution, we assumed the \ce{CO2} dominated desiccated scenario 4, for which all \ce{H2O} is removed from steam scenario 4. We further imposed a weak temperature gradient of 7K/$\ln(p)$ in the radiatively dominated upper atmosphere for the desiccated atmosphere, as observed for Venus, for example, with the Magellan space craft and Pioneer Venus \citep{Jenkins1994}. The opacity sources of the absorbers used in this study are listed in Table~\ref{Tab:Opacity}. For the abiotic cases, we assumed pressure broadening by \ce{CO2} and \ce{H2O}, because 1~bar \ce{N2} is a minor constituent for these cases.

A comparison between the pressure-temperature profile computed with \texttt{1D-TERRA} for the moist 2 scenario compared to scenarios 3 and 4 constructed with a two-component toy model shows interesting similarities (Figure~\ref{fig:MoistGH_Temp}). For example, the upper atmosphere profile of the desiccated scenario 4 and the moist 2 scenario for $p\leq 10^{-2}$~bar are very similar despite different atmosphere compositions. Conversely, the latent heat release of more than 10\% water vapour in the steam-dominated atmospheres (scenarios 3 and 4) shifts the upper atmosphere profiles to higher temperatures compared to the moist scenario 2 with VMR$_{\ce{H2O}}\approx 10^{-4}$ in the upper atmosphere (Figure~\ref{fig:PT_PTERRA}, bottom). This comparison highlights the challenge to distinguish between different atmosphere scenarios of rocky planets in the presence of a high altitude cloud top ($p\leq 10^{-2}$~bar).

\section{Observability of the uninhabitable end-states}
\label{sec: observe_end}

\begin{figure}
   \centering
        \includegraphics[width=0.48\textwidth]{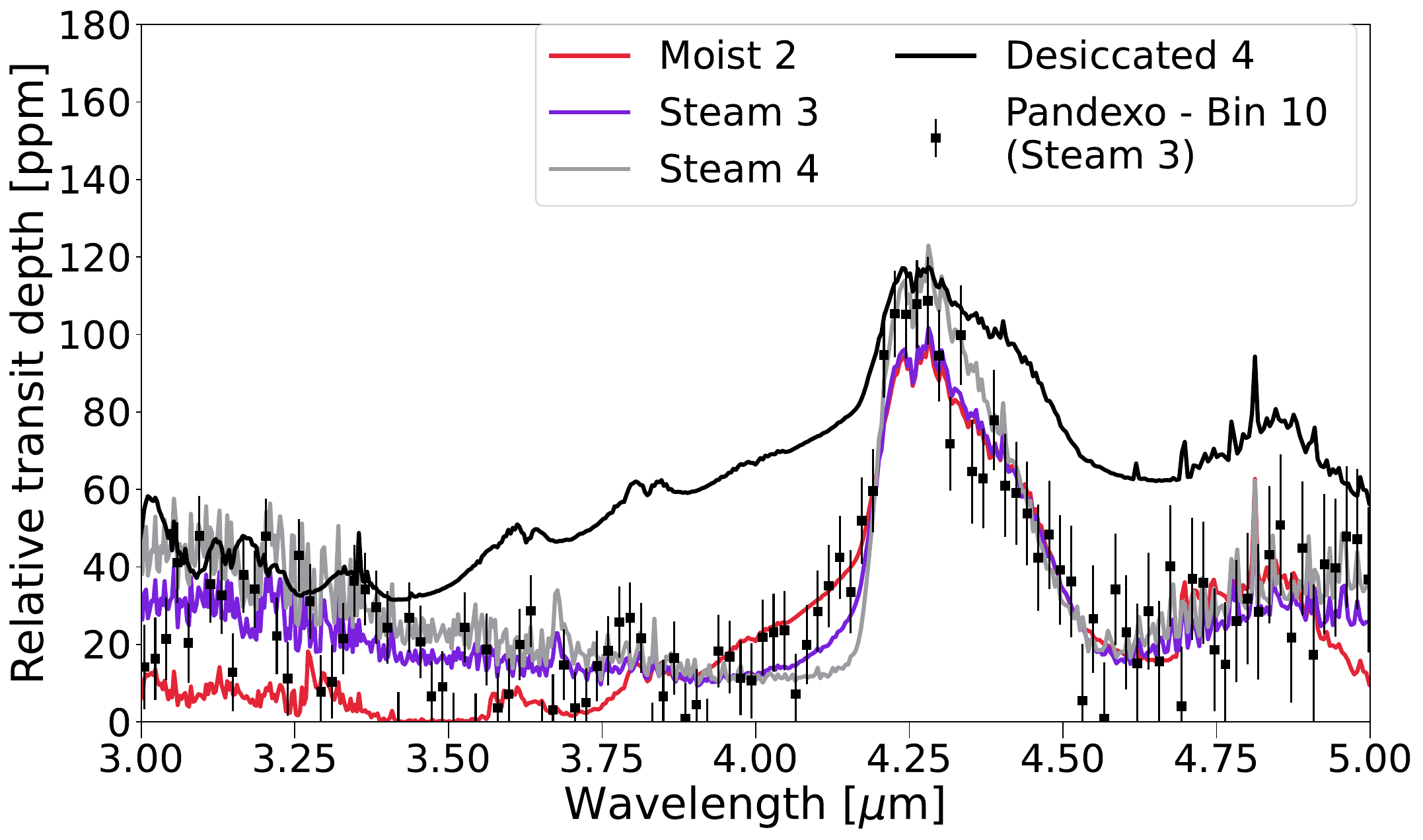}
    \includegraphics[width=0.48\textwidth]{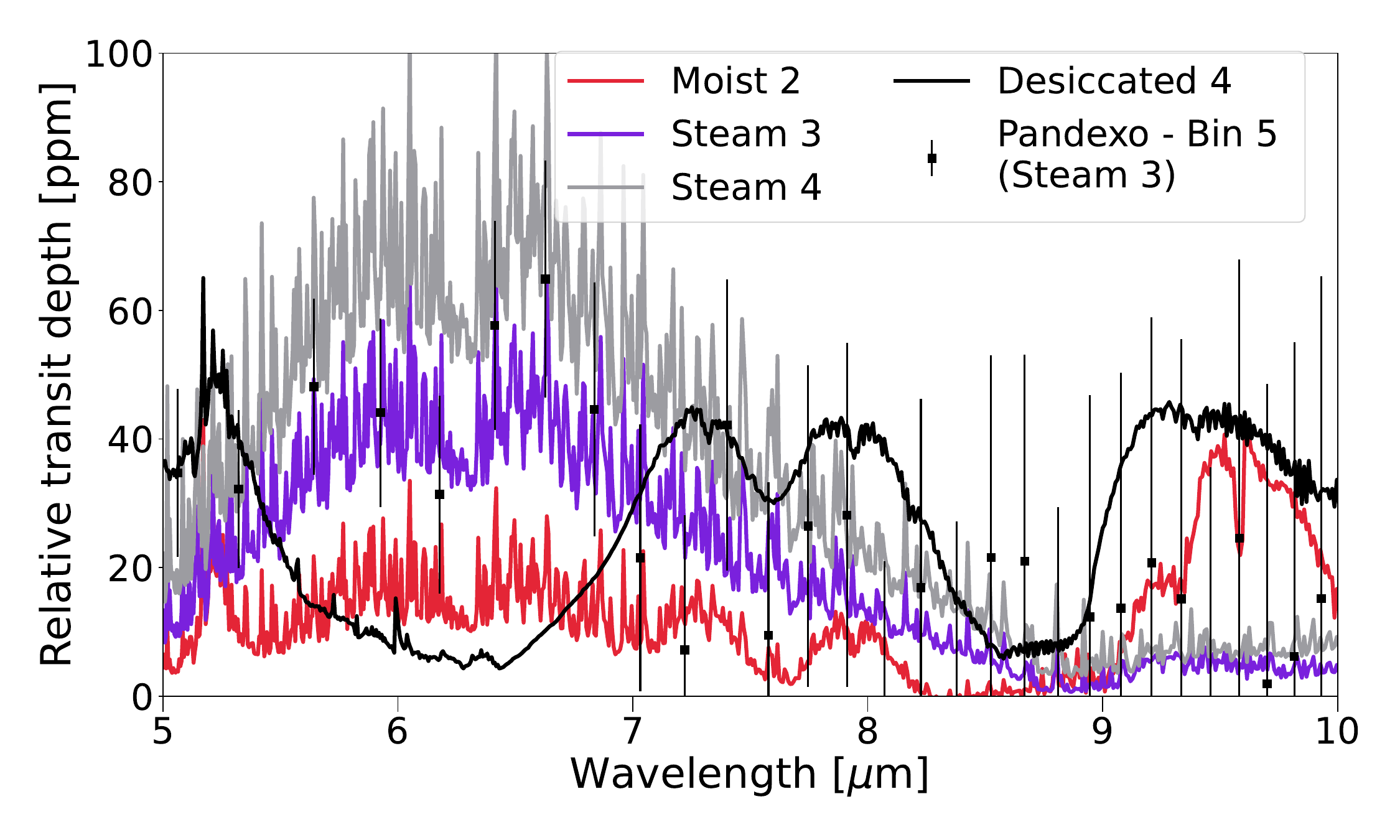}
        \includegraphics[width=0.48\textwidth]{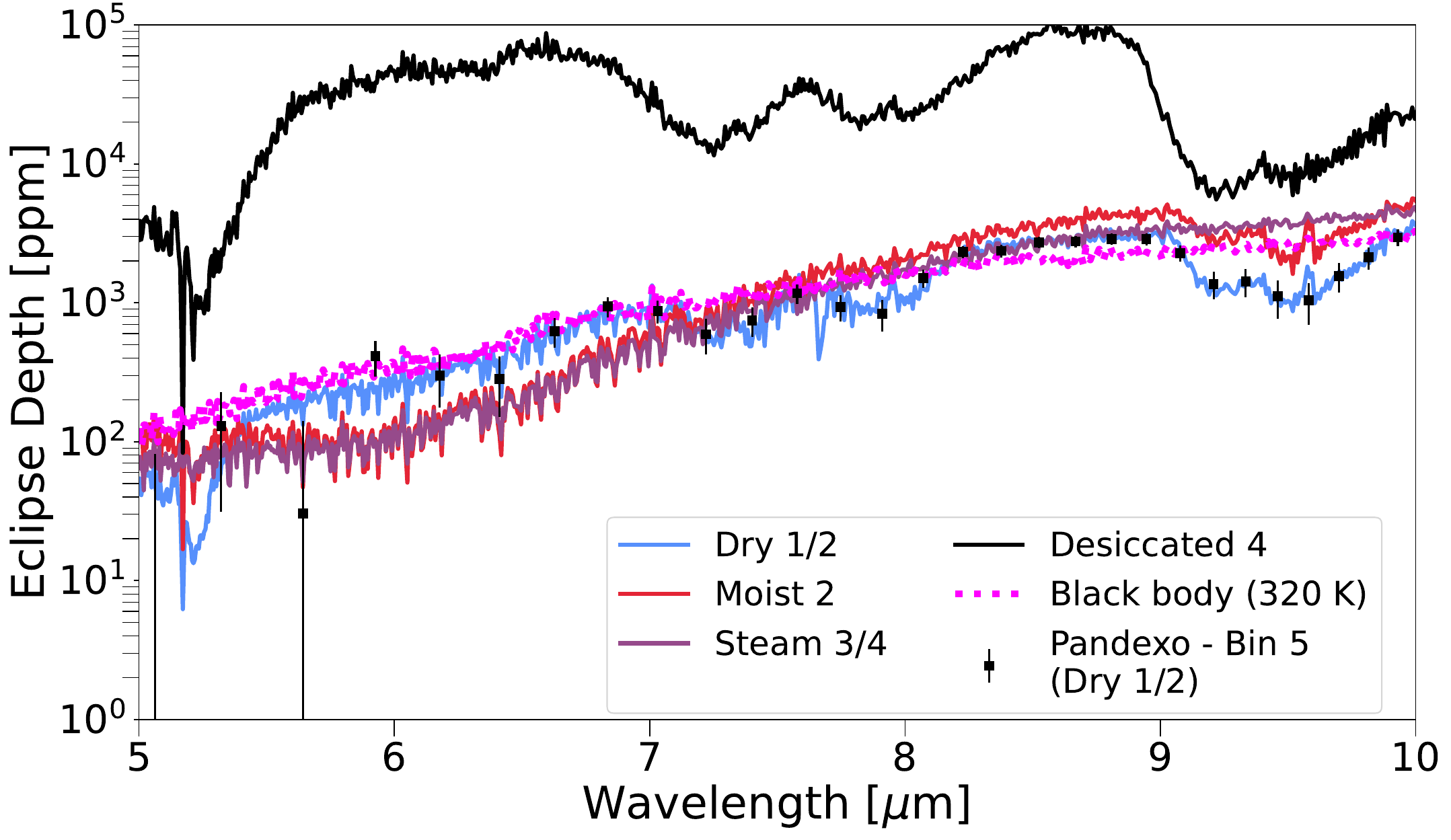}
    \caption{Transmission and eclipse data. The top and centre panels show transmission spectra for moist scenario 2 (red), steam scenarios 3 (purple) and 4 (grey), and the desiccated scenario 4 (black) for the JWST/NIRSpec wavelength range $3-5$~$\mu$m (top) and the JWST/MIRI wavelength range between $5-10$~$\mu$m (centre) with simulated noise (black error bars). We chose the G395M grism setting with 10 pixels per bin (top) and the MIRI LRS setting with 5 pixels per bin (centre). The bottom panel shows eclipse spectra for cloud-free atmospheres for the dry biotic scenarios 1/2 (blue), the moist biotic scenario 2 (red), the abiotic steam scenarios 3/4 (purple), and the desiccated scenario 4 (black) for the JWST/MIRI wavelength range $5-10$~$\mu$m with simulated noise (black error bars), where we combined 5 pixels per bin.}
    \label{fig: End_Transmission}
\end{figure}

In the following, we explore the observational signature of \ce{H2O}-\ce{CO2} dominated and desiccated \ce{CO2} atmosphere scenarios with \texttt{petitRADTRANS} for various JWST observation modes, starting with the cloud-free case. For clarity, we simulated the expected noise from JWST observations with PANDEXO for only one scenario applied to TRAPPIST-1e. We chose the steam 3 scenario for 100 transmission and the dry biotic scenarios for 10 eclipse observations, respectively.

Figure~\ref{fig: End_Transmission} (top) shows that the transition from the moist towards the thick steam-dominated atmosphere regime is mostly heralded by an increase in pressure broadening of the \ce{CO2} absorption feature at 4.3~$\mu$m. For the desiccated \ce{CO2} scenario 4, it is also apparent that this extreme broadening is accompanied by a flattening of the overall transmission spectrum between 3--5~$\mu$m -- even when neglecting the impact of clouds. In the MIRI wavelength range, the water opacities become dominant for the steam atmosphere scenarios between 5--8 $\mu$m (Figure~\ref{fig: End_Transmission}, centre). However, even for the cloud-free steam-dominated scenario 4, the \ce{H2O} amplitude is above 3~$\sigma$ for an accuracy of 20~ppm with 100 JWST transits. The prominent \ce{CO2} spectral features in the desiccated scenario 4 are below 2~$\sigma$.

In the cloud-free case, emission spectra appear to be more promising to differentiate between habitable and uninhabitable climate states. Figure~\ref{fig: End_Transmission} (bottom) demonstrates that \ce{O3} signatures between 9 and 10~$\mu$m could be detectable to an accuracy of 3~$\sigma$ already with 10 JWST eclipse observations. The biosignature of \ce{CH4} at 7.7~$\mu$m is at about 2.5~$\sigma$ significance for 10 observations. Thus, in principle, the signatures of biotic atmospheres could be achievable with relative little time investment with eclipse observations to justify further observations to identify \ce{CH4} and \ce{H2O}. An uninhabitable desiccated \ce{CO2} atmosphere should be easily discernible as well. This conclusion only holds if the atmosphere is cloud-free or at least does not exhibit a thick, continuous cloud coverage, however.

\section{Impact of clouds on the observability of the uninhabitable end states}
\label{sec: Observe_end_cloud}

While the results of the previous section suggest that at least to first order insights about the atmospheric composition could be obtained using JWST transmission and emission spectra, the impact of clouds and of 3D circulation have been neglected, which can affect \ce{O3} concentrations \citep{Carone2018,Chen2019,Chen2021}. For simplicity, we here adopted a continuous, grey cloud deck scenario at $p=0.01$~bar, which represents a thick cumulonimbus cloud for Earth-like atmospheres and the top of the sulphur clouds on Venus \citep[e.g.][]{Cimino1982}. We again simulated the expected noise from JWST observations with PANDEXO for only one scenario applied to TRAPPIST-1e and chose the steam 3 scenario for 100 transmission and the dry biotic scenarios for 10 eclipse observations, respectively.

\begin{figure}
   \centering
        \includegraphics[width=0.48\textwidth]{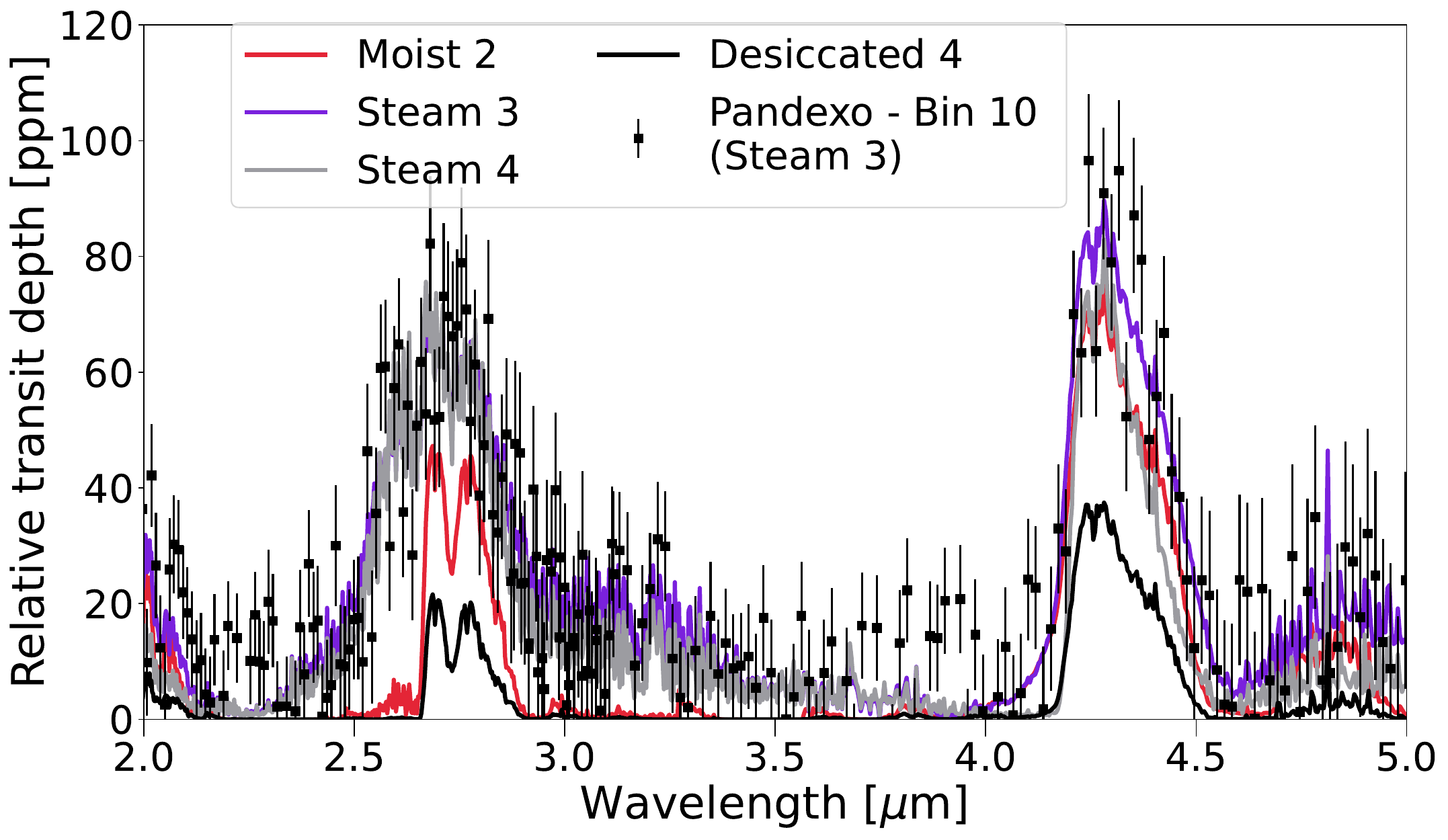}
    \includegraphics[width=0.48\textwidth]{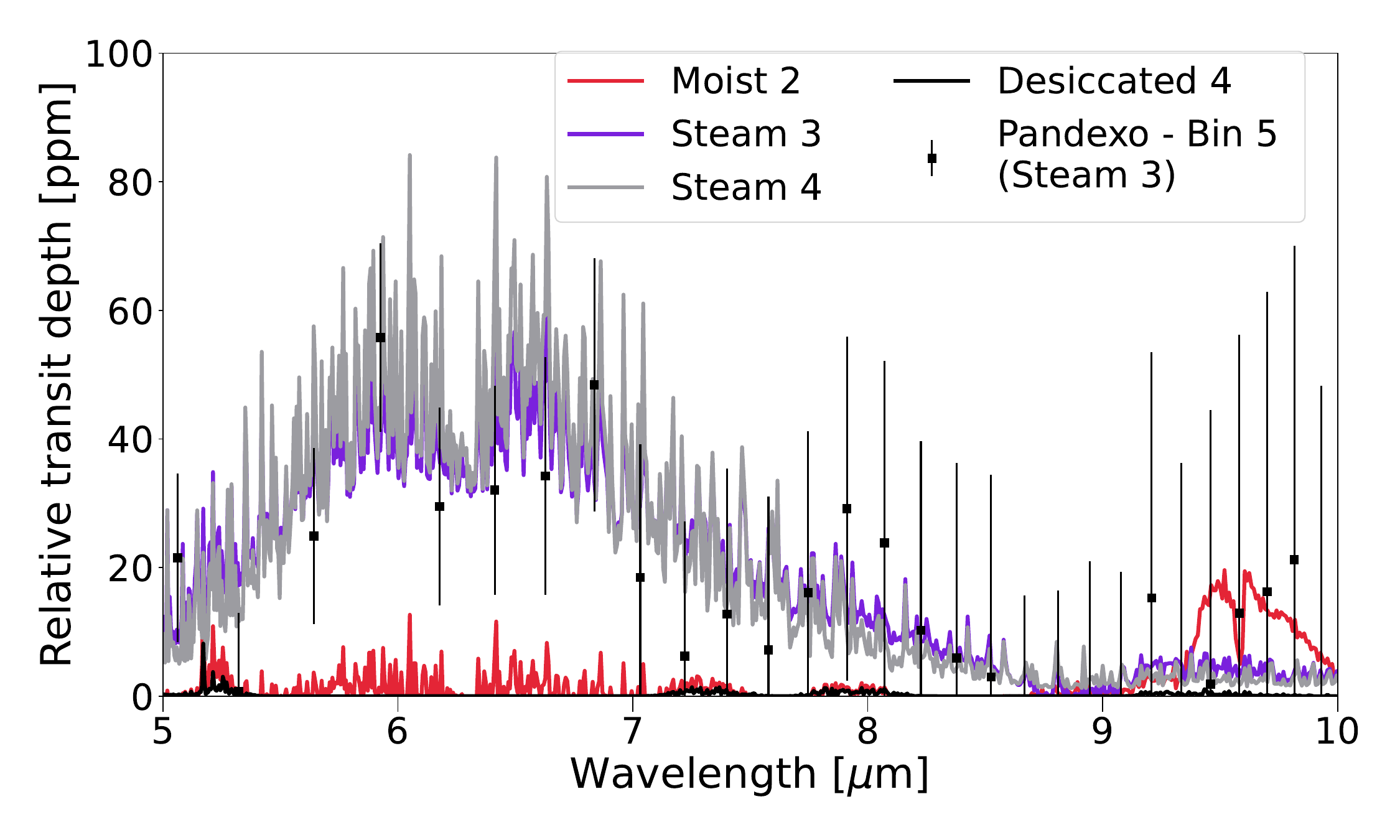}
    \caption{Cloudy transmission spectra. We illustrate the moist, biotic scenario 2 (red), the abiotic steam scenarios 3 (purple) and 4 (grey), and the desiccated scenario 4 for the JWST/NIRSpec wavelength range $3-5$~$\mu$m (top) and the JWST/MIRI wavelength range between $5-10$~$\mu$m (bottom) with simulated noise (black error bars). We chose the G235M (2-3 $\mu$m) and G395M grism setting (3-5 $\mu$m), each with 10 pixels per bin (top), and the MIRI LRS setting with 5 pixels per bin (bottom).}
    \label{fig: End_Transmission_Cloud}
\end{figure}

From the NIRSpec transmission spectrum (Figure~\ref{fig: End_Transmission_Cloud} top), it is evident that a sufficiently high cloud deck reduces all atmospheric features, including the \ce{CO2} feature. Interestingly, a very extended \ce{CO2} dominated atmosphere has the smallest atmospheric features of all explored scenarios. Furthermore, the abiotic steam scenario 4 and the moist biotic scenario 2 would only be distinguishable by the larger water content of the first between 2.5 and 3 micron. Here, however, one would have to ensure that the water feature indeed stems from the atmosphere and not from the host star \citep{Lim2023}. The transmission spectra in the MIRI wavelength range between 5-10 micron would be more illuminating (Figure~\ref{fig: End_Transmission_Cloud} top). Here, the steam scenarios would in principle be differentiable by the \ce{H2O} feature between 5-10 micron. The biotic moist scenario 2 would be identifiable by the \ce{O3} feature at 9.8 $\mu$m. However, the strength of that signal is less than 20~ppm and thus less than 1~$\sigma$ even with 100 transit observations. Only the thick steam atmosphere of the abiotic scenario 4 exhibits a water signature of about 3~$\sigma$. Interestingly, the transmission spectrum between 5 and 10~$\mu$m for a \ce{CO2}-dominated atmosphere of the desiccated scenario 4 is rendered flat by clouds with atmospheric features even below 10~ppm.

\begin{figure}
   \centering
        \includegraphics[width=0.48\textwidth]{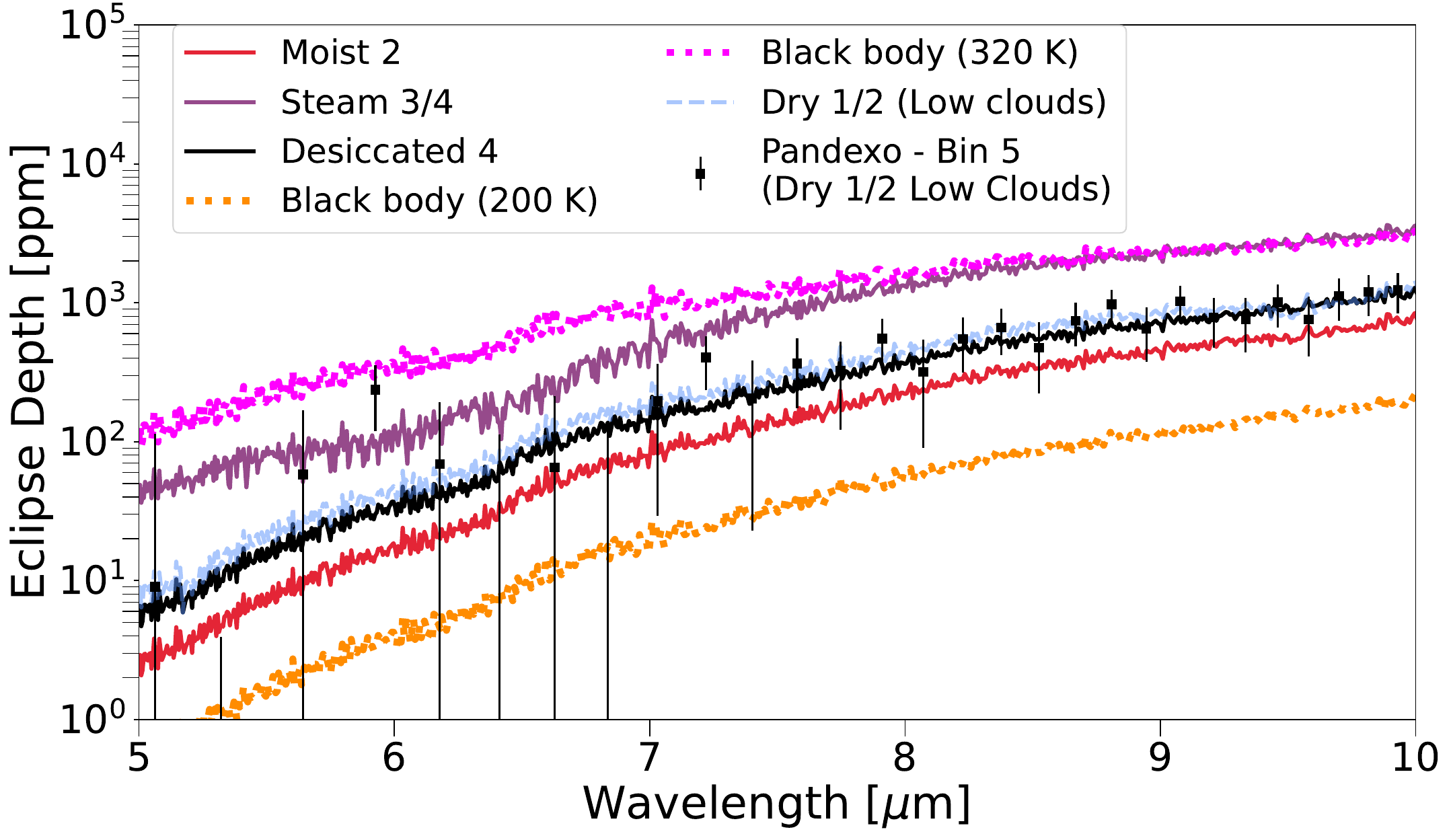}
    \caption{Cloudy eclipse spectra. We show the dry biotic scenarios 1/2 (blue), the moist biotic scenario 2 (red), the abiotic steam scenarios 3/4 (purple), the dry biotic scenario with low clouds (dashed blue), and the desiccated scenario 4 (black) for the JWST/MIRI wavelength range $5-10$~$\mu$m combining 5 pixels per bin (black error bars).}
    \label{fig: End_Eclipse_Cloud}
\end{figure}

The impact of clouds is even more pronounced for measurements focusing on resolving the thermal flux of the planet during the secondary eclipse (Figure~\ref{fig: End_Eclipse_Cloud}). For these measurements, we tested for the dry scenarios 1 and 2 a low cloud top at 0.1~bar. We further simulated JWST observations with PANDEXO for the biotic scenario with low clouds with MIRI LRS, assuming 10 eclipse observations of TRAPPIST-1e. The only remaining noticeable atmospheric feature after a cloud deck was applied is the broad \ce{H2O} feature, which is about 2~$\sigma$ for 10 JWST observations. When the cloud deck was placed high enough ($\leq 10^2$~bar), the differences in the upper atmosphere temperatures become important. We spanned a relatively large range between cool upper atmospheres with 200--250~K for the desiccated \ce{CO2}-dominated and the biotic, habitable atmosphere scenarios (1\&2) and the 320~K warm upper atmosphere for the steam--dominated or runaway-greenhouse atmospheres (scenarios 3 and 4).

For the eclipse measurements, two properties of the upper atmosphere are important. Lower temperatures result in a smaller overall thermal flux from the planet, scaling with $T^4$. At the same time, the low temperature gradient in the upper atmosphere suppresses molecular features like the biosignatures \ce{CH4} and \ce{O3}. Atmospheric characterisation with eclipse measurements will thus only be feasible for runaway greenhouse atmospheres, steam atmospheres, or for atmospheres that are either occasionally cloud-free or exhibit a noncontinuous cloud coverage that allows to resolve the planetary thermal flux from deeper parts of the planetary atmosphere. Otherwise, in the presence of a thick cloud deck, the desiccated scenario 4 with its \ce{CO2}-dominated atmosphere is indistinguishable from biotic habitable scenarios 1 and 2.

Last but not least, we note that the steam scenario 4 has surface temperatures that exceed 1200~K (Figure~\ref{fig:MoistGH_Temp}), which is the melting point of silicate rock \citep{Kasting1988}. In such a case, a release of \ce{SO2} from the surface into the atmosphere could be expected \citep{Janssen2023}.
 
\section{1D-TERRA}
\label{sec: TERRA}

A detailed 1D atmosphere chemistry model sheds more light on the atmosphere composition and observability of the different scenarios. We used the 1D climate-chemistry model 1D-TERRA which is a global mean, stationary, cloud-free, radiative-convective photochemical model extending from the planetary boundary layer, specifically, from a height of 0.5~km above the surface to the lower thermosphere \citep{PavlovKasting02,Rauer11,Gebauer18}. As outlined in \citet{Wunderlich20}, the code has the capability to explore wet and dry climates, motivated by different near-surface assumptions about rain-out in the atmospheric planetary boundary layer (see also Appendix~\ref{sec: PBL}). As a result, the near-surface temperatures may vary due to the strong greenhouse effect of water vapour.

The flexible radiative transfer module REDFOX is described and validated for modern Earth, Venus, and Mars in \citet{Scheucher20}. REDFOX applies the correlated-k distribution method with 128 spectral bands from 100.000 to zero wavenumbers including 20 main absorbers from HITRAN2016 \citep{Gordon17} and 81 absorbers in the FUV to visible from the Max Planck Institute (MPI) Spectral Atlas. Collision-induced absorption coefficients were taken from HITRAN2016 and Rayleigh scattering was included for 8 species. Relative humidity profiles were taken from \citet{ManabeWetherald67}.

The photochemistry module BLACKWOLF was developed for atmospheres dominated by N$_2$, CO$_2$, H$_2$, and H$_2$O. The scheme features 1127 chemical reactions for 128 species. Photochemical reactions for 81 absorbers are considered in 133 bands from 100--850 nm. Rate data were taken form the JPL Report 18 \citep{Burkholder15} and the National Institute of Standards and Technology (NIST) Version 6 \citep{Mallard94}. BLACKWOLF has been validated for modern Earth, Venus and Mars in \citet{Wunderlich20}. Eddy mixing can be calculated flexibly as described in \citet{Wunderlich20}.

A straightforward thermal escape rate for H and O is prescribed in the upper model lid. The lower boundary condition for each chemical species can be set up as a fixed mixing ratio or by source and loss fluxes.

\section{Atmospheric planetary boundary layer}
\label{sec: PBL}
In this work, we connected the outputs of an evolution model of rocky planets, focussing on long-term processes in the mantle, with the surface as the upper boundary and including a simplified atmosphere response, to a detailed 1D atmosphere model (\texttt{1D TERRA}). Coupling these two models, in particular for a case that evolves out of the present-day Earth-like regime, is challenging as it includes the `planetary boundary layer (PBL) problem'. The PBL, extending around 1~km on Earth and 10~km on Mars \citep{Petrosyan2011}, requires detailed treatment of microphysical surface processes such as energy transfer via conduction and convection, turbulence, and water condensation and evaporation, which determines the efficiency of the water cycle and the near-surface latent heat flux. While for present-day Earth, these properties are well measured, they are fare less constrained for other planets. For a study on surface friction within the PBL for tidally-locked Exo-Earths, the reader is referred to \citet{Carone2016}.

Improving the connection between the geophysical and atmospheric models is desirable, yet near-surface conditions are accessible only through in situ measurements, which are currently limited to Solar System planets. Additionally, the conditions of near surface ocean condensation on tidally locked exoplanets \citep{Turbet2023} like TRAPPIST-1e and of moist convection, water vapour transport throughout the free 3D atmosphere, are currently debated \citep{Sergeev2024}. For example, four different 3D climate models for TRAPPIST-1e produce different results in cloud coverage and precipitation under identical irradiation and composition settings \citep{Sergeev2022}. \citet{Godolt2016} point out that for Earth-like atmospheres with increased irradiation, as explored here, the 1D atmosphere model \texttt{TERRA} underestimates the relative humidity compared to 3D climate models when using the Earth-like humidity profiles of \citet{ManabeWetherald67}. To address these uncertainties, we here adopted two extreme water cycle scenarios in \texttt{1D TERRA}.

We assumed that the geophysical model captures the surface conditions, including a water vapour-rich near-surface atmospheric layer at the base of the PBL. \texttt{1D TERRA} then links the surface conditions, mainly above the PBL, to the observable atmosphere, with parameterisations within the PBL to capture parts of the water cycle. The model begins with 1 point at 0.5~km altitude within the PBL. For the dry regimes, we used profiles of \citet{ManabeWetherald67}, assuming most water rains out below the model domain (first 0.5~km), leaving a VMR of water of $10^{-6}$ in the gas phase. The water removed from the gas phase no longer contributes to the greenhouse effect above the rain-out region, resulting in a different temperature at the lower boundary of the pressure-temperature profile at 0.5~km compared to the geophysical model at 0~km height. For the moist regime, only 10\% of the water was assumed to rain-out below the modelling domain, leaving a VMR of water of about $10^{-1}$ in the gas phase above the rain-out region. These two extreme scenarios highlight the importance of the water cycle in the detectability of methane as a potential biosignature.
\end{appendix}
\end{document}